\documentclass[12pt]{article}
\usepackage{fullpage}
\usepackage{times}
\usepackage{isabelle}
\usepackage{isabellesym}
\usepackage{enumerate}
\newcommand{\snip}[4]{\expandafter\newcommand\csname #1\endcsname{#4}}
\snip{decideexistentialdef}{1}{2}{%
\isacommand{definition}\isamarkupfalse%
\ decide{\isacharunderscore}existential\ {\isacharcolon}{\isacharcolon}\ {\isachardoublequoteopen}rat\ poly\ fml\ {\isasymRightarrow}\ bool{\isachardoublequoteclose}\isanewline
\isakeyword{where}\ {\isachardoublequoteopen}decide{\isacharunderscore}existential\ fml\ {\isacharequal}\ {\isacharparenleft}let\ {\isacharparenleft}fml{\isacharunderscore}struct{\isacharcomma}polys{\isacharparenright}\ {\isacharequal}\ convert\ fml\ in\isanewline
\ \ find\ {\isacharparenleft}lookup{\isacharunderscore}sem\ fml{\isacharunderscore}struct{\isacharparenright}\ {\isacharparenleft}find{\isacharunderscore}consistent{\isacharunderscore}signs\ polys{\isacharparenright}\ {\isasymnoteq}\ None{\isacharparenright}{\isachardoublequoteclose}%
}
\snip{decisionprocedurethm}{1}{2}{%
\isacommand{theorem}\isamarkupfalse%
\ decision{\isacharunderscore}procedure{\isacharcolon}\isanewline
\ {\isachardoublequoteopen}{\isacharparenleft}{\isasymforall}x{\isacharcolon}{\isacharcolon}real{\isachardot}\ fml{\isacharunderscore}sem\ fml\ x{\isacharparenright}\ {\isasymlongleftrightarrow}\ decide{\isacharunderscore}universal\ fml{\isachardoublequoteclose}\isanewline
\ {\isachardoublequoteopen}{\isacharparenleft}{\isasymexists}x{\isacharcolon}{\isacharcolon}real{\isachardot}\ fml{\isacharunderscore}sem\ fml\ x{\isacharparenright}\ {\isasymlongleftrightarrow}\ decide{\isacharunderscore}existential\ fml{\isachardoublequoteclose}%
}
\snip{rootsdef}{1}{2}{%
\isacommand{definition}\isamarkupfalse%
\ roots\ {\isacharcolon}{\isacharcolon}\ {\isachardoublequoteopen}real\ poly\ {\isasymRightarrow}\ real\ set{\isachardoublequoteclose}\ \isakeyword{where}\ \isanewline
\ \ {\isachardoublequoteopen}roots\ p\ {\isacharequal}\ {\isacharbraceleft}x{\isachardot}\ poly\ p\ x\ {\isacharequal}\ {\isadigit{0}}{\isacharbraceright}{\isachardoublequoteclose}%
}
\snip{consistentsignsatrootsdef}{1}{2}{%
\isacommand{definition}\isamarkupfalse%
\ consistent{\isacharunderscore}signs{\isacharunderscore}at{\isacharunderscore}roots\ {\isacharcolon}{\isacharcolon}\ {\isachardoublequoteopen}real\ poly\ {\isasymRightarrow}\ real\ poly\ list\ {\isasymRightarrow}\ \isanewline
\ \ rat\ list\ set{\isachardoublequoteclose}\isanewline
\isakeyword{where}\ {\isachardoublequoteopen}consistent{\isacharunderscore}signs{\isacharunderscore}at{\isacharunderscore}roots\ p\ qs\ {\isacharequal}\ {\isacharparenleft}sgn{\isacharunderscore}vec\ qs{\isacharparenright}\ {\isacharbackquote}\ {\isacharparenleft}roots\ p{\isacharparenright}{\isachardoublequoteclose}%
}
\snip{matrixequationthm}{1}{2}{%
\isacommand{theorem}\isamarkupfalse%
\ matrix{\isacharunderscore}equation{\isacharcolon}\isanewline
\isakeyword{assumes}\ {\isachardoublequoteopen}p{\isasymnoteq}{\isadigit{0}}{\isachardoublequoteclose}\isanewline
\isakeyword{assumes}\ {\isachardoublequoteopen}{\isasymAnd}q{\isachardot}\ q\ {\isasymin}\ set\ qs\ {\isasymLongrightarrow}\ coprime\ p\ q{\isachardoublequoteclose}\isanewline
\isakeyword{assumes}\ {\isachardoublequoteopen}distinct\ signs{\isachardoublequoteclose}\isanewline
\isakeyword{assumes}\ {\isachardoublequoteopen}consistent{\isacharunderscore}signs{\isacharunderscore}at{\isacharunderscore}roots\ p\ qs\ {\isasymsubseteq}\ set\ signs{\isachardoublequoteclose}\isanewline
\isakeyword{assumes}\ {\isachardoublequoteopen}{\isasymAnd}l\ i{\isachardot}\ l\ {\isasymin}\ set\ subsets\ {\isasymLongrightarrow}\ i\ {\isasymin}\ set\ l\ {\isasymLongrightarrow}\ i\ {\isacharless}\ length\ qs{\isachardoublequoteclose}\isanewline
\isakeyword{shows}\ {\isachardoublequoteopen}M{\isacharunderscore}mat\ signs\ subsets\ {\isacharasterisk}\isactrlsub v\ w{\isacharunderscore}vec\ p\ qs\ signs\ {\isacharequal}\ v{\isacharunderscore}vec\ p\ qs\ subsets{\isachardoublequoteclose}%
}
\snip{findconsistentsignsatrootsthm}{1}{2}{%
\isacommand{theorem}\isamarkupfalse%
\ find{\isacharunderscore}consistent{\isacharunderscore}signs{\isacharunderscore}at{\isacharunderscore}roots{\isacharcolon}\isanewline
\isakeyword{assumes}\ {\isachardoublequoteopen}p\ {\isasymnoteq}\ {\isadigit{0}}{\isachardoublequoteclose}\isanewline
\isakeyword{assumes}\ {\isachardoublequoteopen}{\isasymAnd}q{\isachardot}\ q\ {\isasymin}\ set\ qs\ {\isasymLongrightarrow}\ coprime\ p\ q{\isachardoublequoteclose}\isanewline
\isakeyword{shows}\ {\isachardoublequoteopen}set\ {\isacharparenleft}find{\isacharunderscore}consistent{\isacharunderscore}signs{\isacharunderscore}at{\isacharunderscore}roots\ p\ qs{\isacharparenright}\ {\isacharequal}\ \isanewline
\ \ consistent{\isacharunderscore}signs{\isacharunderscore}at{\isacharunderscore}roots\ p\ qs{\isachardoublequoteclose}%
}
\snip{calcdatadef}{1}{2}{%
\isacommand{fun}\isamarkupfalse%
\ calc{\isacharunderscore}data\ {\isacharcolon}{\isacharcolon}\ {\isachardoublequoteopen}real\ poly\ {\isasymRightarrow}\ real\ poly\ list\ {\isasymRightarrow}\ \isanewline
\ \ {\isacharparenleft}rat\ mat\ {\isasymtimes}\ {\isacharparenleft}nat\ list\ list\ {\isasymtimes}\ rat\ list\ list{\isacharparenright}{\isacharparenright}{\isachardoublequoteclose}\isanewline
\isakeyword{where}\ {\isachardoublequoteopen}calc{\isacharunderscore}data\ p\ qs\ {\isacharequal}\ {\isacharparenleft}let\ len\ {\isacharequal}\ length\ qs\ in\isanewline
if\ len\ {\isacharequal}\ {\isadigit{0}}\ then\isanewline
\ \ {\isacharparenleft}{\isasymlambda}{\isacharparenleft}a{\isacharcomma}b{\isacharcomma}c{\isacharparenright}{\isachardot}{\isacharparenleft}a{\isacharcomma}b{\isacharcomma}map\ {\isacharparenleft}drop\ {\isadigit{1}}{\isacharparenright}\ c{\isacharparenright}{\isacharparenright}\ {\isacharparenleft}reduce{\isacharunderscore}system\ p\ {\isacharparenleft}{\isacharbrackleft}{\isadigit{1}}{\isacharbrackright}{\isacharcomma}base{\isacharunderscore}case{\isacharunderscore}info{\isacharparenright}{\isacharparenright}\isanewline
else\ if\ len\ {\isasymle}\ {\isadigit{1}}\ then\ reduce{\isacharunderscore}system\ p\ {\isacharparenleft}qs{\isacharcomma}base{\isacharunderscore}case{\isacharunderscore}info{\isacharparenright}\isanewline
else\ {\isacharparenleft}let\ qs{\isadigit{1}}\ {\isacharequal}\ take\ {\isacharparenleft}len\ div\ {\isadigit{2}}{\isacharparenright}\ qs{\isacharsemicolon}\ left\ {\isacharequal}\ calc{\isacharunderscore}data\ p\ qs{\isadigit{1}}{\isacharsemicolon}\isanewline
\ \ \ \ \ \ \ \ \ \ qs{\isadigit{2}}\ {\isacharequal}\ drop\ {\isacharparenleft}len\ div\ {\isadigit{2}}{\isacharparenright}\ qs{\isacharsemicolon}\ right\ {\isacharequal}\ calc{\isacharunderscore}data\ p\ qs{\isadigit{2}}\ in\isanewline
\ \ \ \ \ \ \ \ reduce{\isacharunderscore}system\ p\ {\isacharparenleft}combine{\isacharunderscore}systems\ p\ {\isacharparenleft}qs{\isadigit{1}}{\isacharcomma}left{\isacharparenright}\ {\isacharparenleft}qs{\isadigit{2}}{\isacharcomma}right{\isacharparenright}{\isacharparenright}{\isacharparenright}{\isacharparenright}{\isachardoublequoteclose}%
}
\snip{findconsistentsignsatrootsdef}{1}{2}{%
\isacommand{definition}\isamarkupfalse%
\ find{\isacharunderscore}consistent{\isacharunderscore}signs{\isacharunderscore}at{\isacharunderscore}roots\ {\isacharcolon}{\isacharcolon}\isanewline
\ {\isachardoublequoteopen}real\ poly\ {\isasymRightarrow}\ real\ poly\ list\ {\isasymRightarrow}\ rat\ list\ list{\isachardoublequoteclose}\isanewline
\ \isakeyword{where}\ {\isachardoublequoteopen}find{\isacharunderscore}consistent{\isacharunderscore}signs{\isacharunderscore}at{\isacharunderscore}roots\ p\ qs\ {\isacharequal}\ \isanewline
\ \ {\isacharparenleft}let\ {\isacharparenleft}M{\isacharcomma}S{\isacharcomma}{\isasymSigma}{\isacharparenright}\ {\isacharequal}\ calc{\isacharunderscore}data\ p\ qs\ in\ {\isasymSigma}{\isacharparenright}{\isachardoublequoteclose}%
}
\snip{kroneckerproductdef}{1}{2}{%
\isacommand{definition}\isamarkupfalse%
\ kronecker{\isacharunderscore}product\ {\isacharcolon}{\isacharcolon}\ {\isachardoublequoteopen}{\isacharprime}a\ {\isacharcolon}{\isacharcolon}\ ring\ mat\ {\isasymRightarrow}\ {\isacharprime}a\ mat\ {\isasymRightarrow}\ {\isacharprime}a\ mat{\isachardoublequoteclose}\isanewline
\isakeyword{where}\ {\isachardoublequoteopen}kronecker{\isacharunderscore}product\ A\ B\ {\isacharequal}\ {\isacharparenleft}\isanewline
let\ ra\ {\isacharequal}\ dim{\isacharunderscore}row\ A{\isacharsemicolon}\ ca\ {\isacharequal}\ dim{\isacharunderscore}col\ A{\isacharsemicolon}\ rb\ {\isacharequal}\ dim{\isacharunderscore}row\ B{\isacharsemicolon}\ cb\ {\isacharequal}\ dim{\isacharunderscore}col\ B\ in\isanewline
\ \ mat\ {\isacharparenleft}ra\ {\isacharasterisk}\ rb{\isacharparenright}\ {\isacharparenleft}ca\ {\isacharasterisk}\ cb{\isacharparenright}\isanewline
\ \ \ \ {\isacharparenleft}{\isasymlambda}{\isacharparenleft}i{\isacharcomma}j{\isacharparenright}{\isachardot}\ A\ {\isachardollar}{\isachardollar}\ {\isacharparenleft}i\ div\ rb{\isacharcomma}\ j\ div\ cb{\isacharparenright}\ {\isacharasterisk}\ B\ {\isachardollar}{\isachardollar}\ {\isacharparenleft}i\ mod\ rb{\isacharcomma}\ j\ mod\ cb{\isacharparenright}{\isacharparenright}{\isacharparenright}{\isachardoublequoteclose}%
}
\snip{rowstokeepdef}{1}{2}{%
\isacommand{definition}\isamarkupfalse%
\ rows{\isacharunderscore}to{\isacharunderscore}keep{\isacharcolon}{\isacharcolon}\ {\isachardoublequoteopen}{\isacharparenleft}{\isacharprime}a{\isacharcolon}{\isacharcolon}field{\isacharparenright}\ mat\ {\isasymRightarrow}\ nat\ list{\isachardoublequoteclose}\isanewline
\ \ \isakeyword{where}\ {\isachardoublequoteopen}rows{\isacharunderscore}to{\isacharunderscore}keep\ A\ {\isacharequal}\ map\ snd\ {\isacharparenleft}pivot{\isacharunderscore}positions\ \isanewline
\ \ {\isacharparenleft}gauss{\isacharunderscore}jordan{\isacharunderscore}single\ {\isacharparenleft}A\isactrlsup T{\isacharparenright}{\isacharparenright}{\isacharparenright}{\isachardoublequoteclose}%
}
\snip{rowstokeepranklem}{1}{2}{%
\isacommand{lemma}\isamarkupfalse%
\ rows{\isacharunderscore}to{\isacharunderscore}keep{\isacharunderscore}rank{\isacharcolon}\isanewline
\isakeyword{assumes}\ {\isachardoublequoteopen}dim{\isacharunderscore}col\ A\ {\isasymle}\ dim{\isacharunderscore}row\ A{\isachardoublequoteclose}\isanewline
\isakeyword{shows}\ {\isachardoublequoteopen}vec{\isacharunderscore}space{\isachardot}rank\ {\isacharparenleft}length\ {\isacharparenleft}rows{\isacharunderscore}to{\isacharunderscore}keep\ A{\isacharparenright}{\isacharparenright}\ {\isacharparenleft}take{\isacharunderscore}rows\ A\ \isanewline
\ \ {\isacharparenleft}rows{\isacharunderscore}to{\isacharunderscore}keep\ A{\isacharparenright}{\isacharparenright}\ {\isacharequal}\ vec{\isacharunderscore}space{\isachardot}rank\ {\isacharparenleft}dim{\isacharunderscore}row\ A{\isacharparenright}\ A{\isachardoublequoteclose}%
}
\snip{transposeranklem}{1}{2}{%
\isacommand{lemma}\isamarkupfalse%
\ transpose{\isacharunderscore}rank{\isacharcolon}\isanewline
\isakeyword{shows}\ {\isachardoublequoteopen}vec{\isacharunderscore}space{\isachardot}rank\ {\isacharparenleft}dim{\isacharunderscore}row\ A{\isacharparenright}\ A\ {\isacharequal}\ vec{\isacharunderscore}space{\isachardot}rank\ {\isacharparenleft}dim{\isacharunderscore}col\ A{\isacharparenright}\ {\isacharparenleft}A\isactrlsup T{\isacharparenright}{\isachardoublequoteclose}%
}

\usepackage[bookmarks=false]{hyperref}

\usepackage{doi}

\usepackage{hyphenat}

\usepackage{amsthm,amsmath,amssymb}
\usepackage[dvipsnames]{xcolor}
\usepackage{proof}
\usepackage{graphicx}
\usepackage{relsize}
 \usepackage{float}
\usepackage{exscale}
\usepackage{ebproof}
\usepackage{hyperref}

\usepackage{verbatim}
\usepackage{stmaryrd}
\usepackage{multicol}
\usepackage{appendix}
\usepackage{etex}
\usepackage{etoolbox}
\usepackage{xspace}

\usepackage{amsthm}

\theoremstyle{plain}
\newtheorem{theorem}{Theorem}

\theoremstyle{definition}
\newtheorem{definition}{Definition}
\theoremstyle{remark}
\newtheorem{remark}{Remark}

\usepackage{booktabs}
\usepackage{bm}

\usepackage{math}
\usepackage[prefixflatinterpret,bracketmodalinterpret,fixformat,silentconst,sidenotecalculus,longseqcontext,seqinsist]{logic}

\usepackage{prettyref}
\newcommand{\rref}[2][]{\prettyref{#2}}
\newrefformat{sec}{Section\,\ref{#1}}
\newrefformat{subsec}{Section\,\ref{#1}}
\newrefformat{def}{Def.\,\ref{#1}}
\newrefformat{thm}{Theorem\,\ref{#1}}
\newrefformat{prop}{Proposition\,\ref{#1}}
\newrefformat{rem}{Remark\,\ref{#1}}
\newrefformat{lem}{Lemma\,\ref{#1}}
\newrefformat{cor}{Corollary\,\ref{#1}}
\newrefformat{ex}{Example\,\ref{#1}}
\newrefformat{cex}{Counterexample\,\ref{#1}}
\newrefformat{tab}{Table\,\ref{#1}}
\newrefformat{fig}{Fig.\,\ref{#1}}
\newrefformat{case}{case\,\ref{#1}}
\newrefformat{foot}{Footnote\,\ref{#1}}

\newcommand{\bebecomes}{\mathrel{::=}}
\newcommand{\alternative}{~|~}
\newcommand{\Isabelle}{Isabelle/HOL\xspace}

\usepackage{tikz}
\usetikzlibrary{arrows}

\newcommand*\circled[1]{\tikz[baseline=(char.base)]{
            \node[shape=circle,draw,inner sep=1pt] (char) {#1};}}

\usepackage{fancyhdr}
\title{A Verified Decision Procedure for Univariate Real Arithmetic with the BKR Algorithm}
\author{Katherine Cordwell \and Yong Kiam Tan \and
Andr\'e Platzer\thanks{Computer Science Department, Carnegie Mellon University, Pittsburgh, USA\newline $\{$kcordwel,yongkiat,aplatzer$\}$@cs.cmu.edu}}
\date{}

\usepackage{ifpdf}
\ifpdf
\fi

\begin{document}
\maketitle
\allowdisplaybreaks
\thispagestyle{empty}
\begin{abstract}
We formalize the univariate fragment of Ben-Or, Kozen, and Reif's (BKR) decision procedure for first-order real arithmetic in Isabelle/HOL.
BKR's algorithm has good potential for parallelism and was designed to be used in practice.
Its key insight is a clever recursive procedure that computes the set of all consistent sign assignments for an input set of univariate polynomials while carefully managing intermediate steps to avoid exponential blowup from naively enumerating all possible sign assignments (this insight is fundamental for both the univariate case and the general case).
Our proof combines ideas from BKR and a follow-up work by Renegar that are well-suited for formalization.
The resulting proof outline allows us to build substantially on Isabelle/HOL's libraries for algebra, analysis, and matrices.
Our main extensions to existing libraries are also detailed.
\end{abstract}

\section{Introduction}
\label{sec:Introduction}
Formally verified arithmetic has important applications in formalized mathematics and rigorous engineering domains.
For example, real arithmetic questions (first-order formulas in the \textit{theory of real closed fields}) often arise as part of formal proofs for safety-critical cyber-physical systems (CPS)~\cite{Platzer18},
the formal proof of the Kepler conjecture involves the verification of more than $23,000$ real inequalities~\cite{hales_2017}, and the verification of floating-point algorithms also involves real arithmetic reasoning~\cite{DBLP:conf/sfm/Harrison06}.
Some real arithmetic questions involve $\forall$ and $\exists$ quantifiers; these \textit{quantified} real arithmetic questions arise in, e.g., CPS proofs, geometric theorem proving, and stability analysis of models of biological systems \cite{DBLP:journals/mics/Sturm17}.

\textit{Quantifier elimination (QE)} is the process by which a quantified formula is transformed into a logically equivalent quantifier-free formula.
Tarski famously proved that the theory of first-order real arithmetic (FOL$_{\mathbb{R}}$) admits QE; FOL$_{\mathbb{R}}$ validity and satisfiability are therefore decidable by QE and evaluation~\cite{Tarski}.
Thus, \emph{in theory}, all it takes to rigorously answer \emph{any} real arithmetic question is to verify a QE procedure for FOL$_{\mathbb{R}}$.
However, \emph{in practice}, QE algorithms for FOL$_{\mathbb{R}}$ are complicated and the fastest known QE algorithm, \textit{cylindrical algebraic decomposition (CAD)} \cite{Collins} is, in the worst case, doubly exponential in the number of variables.
The multivariate CAD algorithm is highly complicated and has yet to be fully formally verified in a theorem prover~\cite{li2019deciding}, although various specialized approaches have been used to successfully tackle restricted subsets of real arithmetic questions in proof assistants, e.g., quantifier elimination for linear real arithmetic~\cite{DBLP:journals/jar/Nipkow10}, sum-of-squares witnesses~\cite{DBLP:conf/tphol/Harrison07} or real Nullstellensatz witnesses~\cite{DBLP:conf/cade/PlatzerQR09} for the universal fragment, and interval arithmetic when the quantified variables range over bounded domains~\cite{hoelzl09realineequalities,pittir16721}.

There are few general-purpose formally verified decision procedures for FOL$_{\mathbb{R}}$.
Mahboubi and Cohen \cite{AssiaQE} formally verified an algorithm for QE based on Tarski's proof but their formalization is primarily a theoretical decidability result \cite[Section 1]{AssiaQE} owing to the non-elementary complexity of Tarski's algorithm.
The proof-producing procedure by McLaughlin and Harrison \cite{harrison} can solve a number of small multivariate examples but suffers similarly from the complexity of the underlying Cohen-H\"{o}rmander procedure.
The situation for \textit{univariate real arithmetic} (i.e., problems that involve only a single variable) is better.
In \Isabelle, Li, Passmore, and Paulson \cite{li2019deciding} formalized an efficient univariate decision procedure based on univariate CAD.
There are additionally some univariate decision procedures in PVS, including $\mathtt{hutch}$~\cite{NASAHutch} (based on CAD) and $\mathtt{tarski}$~\cite{NASATarski} (based on the Sturm-Tarski theorem).

This paper adds to the latter body of work by formalizing the univariate case of Ben-Or, Kozen, and Reif's (BKR) decision procedure \cite{DBLP:journals/jcss/Ben-OrKR86}  in Isabelle/HOL \cite{DBLP:books/sp/NipkowPW02, DBLP:journals/jar/Paulson89}.
Our formalization of univariate BKR is ${\approx}7000$ lines \cite{BKR_AFP}.
Our main contributions are:
\begin{itemize}
\item In~\rref{sec:mathematics}, we present an algorithmic blueprint for implementing BKR's procedure that blends insights from Renegar's \cite{DBLP:journals/jsc/Renegar92b} later variation of BKR.
Compared to the original abstract presentations~\cite{DBLP:journals/jcss/Ben-OrKR86,DBLP:journals/jsc/Renegar92b}, our blueprint is phrased concretely in terms of matrix operations which facilitates its implementation and identifies its correctness properties.
\item Our blueprint is designed for formalization by judiciously combining and fleshing out BKR's and Renegar's proofs. In \rref{sec:formalization}, we outline key aspects of our proof, its use of existing \Isabelle libraries, and our contributions to those libraries.
\end{itemize}

It is desirable to have a variety of formally verified decision procedures for arithmetic since different strategies can have different efficiency tradeoffs on different classes of problems~\cite{NASATechRept,DBLP:conf/cade/PlatzerQR09}.
For example, in PVS, $\mathtt{hutch}$ is usually significantly faster than $\mathtt{tarski}$ \cite{NASAHutch} but there are a number of adversarial problems for $\mathtt{hutch}$ on which $\mathtt{tarski}$ performs better \cite{NASATechRept}.
BKR has a fundamentally different working principle than CAD; like the Cohen-H\"ormander procedure, it represents roots and sign-invariant regions abstractly, instead of via computationally expensive, real algebraic numbers required in CAD.
Further, unlike Cohen-H\"ormander, BKR was designed to be used in practice: when its inherent parallelism is exploited, an optimized version of univariate BKR is an NC algorithm (that is, it runs in parallel polylogarithmic time).
Our formalization is not yet optimized and parallelized, so we do not yet achieve such efficiency.
However, we do export our Isabelle/HOL formalization to Standard ML (SML) and are able to solve some examples with the exported code (\rref{sec:code}).

Additionally, our formalization is a significant stepping stone towards the multivariate case, which builds inductively on the univariate case.
We give some (informal) mathematical intuition for multivariate BKR in Appendix \ref{sec:multivar}---since multivariate BKR seems to rely fairly directly on the univariate version, we hope that it will be significantly easier to formally verify than multivariate CAD, which is highly complicated.
However, it is unlikely that multivariate BKR will be as efficient as CAD in the average case.
While BKR states that their multivariate algorithm is computable in parallel exponential time (or in NC for fixed dimension), Canny later found an error in BKR's analysis of the multivariate case~\cite{1993Improved}, which highlights the subtlety of the algorithm and the role for formal verification.
Notwithstanding this, multivariate BKR is almost certain to outperform methods such as Tarski's algorithm and Cohen-H\"{o}rmander and can supplement an eventual formalization of multivariate CAD.

Our formalization is available on the Archive of Formal Proofs (AFP) \cite{BKR_AFP}.
%

\section{Mathematical Underpinnings}\label{sec:mathematics}
This section provides an outline of our decision procedure for univariate real arithmetic and its verification in \Isabelle \cite{DBLP:books/sp/NipkowPW02}.
The goal is to provide an accessible mathematical blueprint that explains our construction and its blend of ideas from BKR~\cite{DBLP:journals/jcss/Ben-OrKR86} and Renegar~\cite{DBLP:journals/jsc/Renegar92b}; in-depth technical discussion of the formal proofs is largely deferred to~\rref{sec:formalization}.
Our procedure starts with two transformation steps (Sections~\ref{sec:QEtoSD} and~\ref{sec:settingup}) that simplify an input decision problem into a so-called restricted sign determination format.
An algorithm for the latter problem is then presented in~\rref{sec:keystep}.
Throughout this paper, unless explicitly specified, we are working with \textit{univariate} polynomials, which we assume to have variable $x$.
Our decision procedure works for polynomials with rational coefficients (\isa{rat poly} in Isabelle), though some lemmas are proved more generally for univariate polynomials with real coefficients (\isa{real poly} in Isabelle).

\subsection{From Univariate Problems to Sign Determination}\label{sec:QEtoSD}
Formulas of \emph{univariate real arithmetic} are generated by the following grammar, where $p$ is a univariate polynomial with rational coefficients:
\[
	\phi,\psi~\bebecomes~p > 0 \alternative p \geq 0 \alternative p = 0 \alternative \phi \lor \psi \alternative \phi \land \psi
\]
In \Isabelle, we define this grammar in \isa{fml}, which is our type for formulas.

For formula $\phi$, the \emph{universal} decision problem is to decide if $\phi$ is true for \emph{all} real values of $x$, i.e., validity of the quantified formula $\lforall{x}{\phi}$.
The \emph{existential} decision problem is to decide if $\phi$ is true for \emph{some} real value of $x$, i.e., validity of the quantified formula $\lexists{x}{\phi}$.
For example, a decision procedure should return false for formula~\rref{eq:QEex1-1} and true for formula~\rref{eq:QEex1-2} below (left).

\noindent\begin{minipage}[t]{0.45\textwidth}%
\vspace{-5mm}
\begin{align}%
\lforall{x}{(x^2 - 2= 0 \land 3x > 0)} \label{eq:QEex1-1} \\
\lexists{x}{(x^2 - 2= 0 \land 3x > 0)} \label{eq:QEex1-2}
\end{align}%
\end{minipage}%
\hfill\vline\hfill%
\begin{minipage}[t]{0.52\textwidth}%
\vspace{-5mm}
\begin{align*}%
&\text{Formula Structure:} && \circled{A} = 0 \land \circled{B} > 0\\
&\text{Polynomials:} && \circled{A}: x^2 - 2,\quad \circled{B}: 3x
\end{align*}%
\end{minipage}
\vspace{2mm}

The first observation is that both univariate decision problems can be transformed to the problem of finding the set of \emph{consistent sign assignments} (also known as realizable sign assignments~\cite[Definition 2.34]{algRAG}) of the set of polynomials appearing in the formula $\phi$.

\begin{definition}
A \textbf{sign assignment} for a set $G$ of polynomials is a mapping $\sigma$ that assigns each $g \in G$ to either $+1$, $-1$, or $0$.
A sign assignment $\sigma$ for $G$ is \textbf{consistent} if there exists an $x \in \reals$ where, for all $g \in G$, the sign of $g(x)$ matches the sign of $\sigma(g)$.
\end{definition}

For the polynomials $x^2 - 2$ and $3x$ appearing in formulas~\rref{eq:QEex1-1} and~\rref{eq:QEex1-2}, the set of all consistent sign assignments (written as ordered pairs) is:
\[ \{ (+1, -1),~(0, -1),~(-1, -1),~(-1, 0),~(-1, +1),~(0, +1),~(+1, +1) \} \]

Formula~\rref{eq:QEex1-1} is not valid because consistency of sign assignment $(0,-1)$ implies there exists a real value $x \in \reals$ such that conjunct $x^2-2=0$ is satisfied but not $3x > 0$.
Conversely, formula~\rref{eq:QEex1-2} is valid because the consistent sign assignment $(0,+1)$ demonstrates the existence of an $x \in \reals$ satisfying $x^2-2=0$ and $3x > 0$.
The truth-value of formula $\phi$ at a given sign assignment is computed by evaluating the formula after replacing all of its polynomials by their respective assigned signs.
For example, for the sign assignment $(0,-1)$, replacing $\circled{A}$ by $0$ and $\circled{B}$ by $-1$ in the formula structure underlying~\rref{eq:QEex1-1} and~\rref{eq:QEex1-2} shown above (right) yields $0=0 \land -1 > 0$, which evaluates to false.
Validity of $\lforall{x}{\phi}$ is decided by checking that $\phi$ evaluates to true at \emph{each} of its consistent sign assignments.
Similarly, validity of $\lexists{x}{\phi}$ is decided by checking that $\phi$ evaluates to true at \emph{at least one} consistent sign assignment.

Our top-level formalized algorithms are called \isa{decide\isacharunderscore universal} and \isa{decide\isacharunderscore existential}, both with type \isa{rat poly fml $\Rightarrow$ bool}.
The definition of \isa{decide\isacharunderscore existential} is as follows (the omitted definition of \isa{decide\isacharunderscore universal} is similar):
\begin{isabelle}
\decideexistentialdef
\end{isabelle}

Here, \isa{convert} extracts the list of constituent polynomials \isa{polys} from the input formula \isa{fml} along with the formula structure \isa{fml\isacharunderscore struct}, \isa{find\isacharunderscore consistent\isacharunderscore signs} returns the list of all consistent sign assignments \isa{conds} for \isa{polys}, and \isa{find} checks that predicate \isa{lookup\isacharunderscore sem fml\isacharunderscore struct} is true at one of those sign assignments.
Given a sign assignment $\sigma$, \isa{lookup\isacharunderscore sem fml\isacharunderscore struct $\sigma$} evaluates the truth value of \isa{fml} at $\sigma$ by recursively evaluating the truth of its subformulas after replacing polynomials by their sign according to $\sigma$ using the formula structure \isa{fml\isacharunderscore struct}.
Thus, \isa{decide\isacharunderscore existential} returns true iff \isa{fml} evaluates to true for at least one of the consistent sign assignments of its constituent polynomials.

The correctness theorem for \isa{decide\isacharunderscore universal} and \isa{decide\isacharunderscore existential} is shown below, where \isa{fml\isacharunderscore sem fml x} evaluates the truth of formula \isa{fml} at the real value \isa{x}.
\begin{isabelle}
\decisionprocedurethm
\end{isabelle}

This theorem depends crucially on \isa{find\isacharunderscore consistent\isacharunderscore signs} correctly finding \emph{all} consistent sign assignments for \isa{polys}, i.e., solving the sign determination problem.

\subsection {From Sign Determination to Restricted Sign Determination}\label{sec:settingup}
The next step restricts the sign determination problem to the following more concrete format: Find all consistent sign assignments $\sigma$ for a set of polynomials $q_1, \dots, q_n$ at the roots of a nonzero polynomial $p$, i.e., the signs of $q_1(x),\dots,q_n(x)$ that occur at the (finitely many) real values $x \in \reals$ with $p(x)=0$.
The key insight of BKR is that this restricted problem can be solved efficiently (in parallel) using purely algebraic tools (\rref{sec:keystep}).
Following BKR's procedure, we also normalize the $q_i$'s to be coprime with (i.e. share no common factors with) $p$, which simplifies the subsequent construction for the key step and its formal proof.

\begin{remark}
The normalization of $q_i$'s to be coprime with $p$ can be avoided using a slightly more intricate construction due to Renegar~\cite{DBLP:journals/jsc/Renegar92b}.
We have also formalized this construction but omit full details in this paper as the formalization was completed after acceptance for publication. Its overall structure is quite similar to~\rref{sec:keystep}, and it is available in the AFP alongside our formalization of BKR \cite{BKR_AFP}.
\end{remark}


Consider as input a set of polynomials (with rational coefficients) $G = \{g_1, \dots, g_k\}$ for which we need to find all consistent sign assignments.
The transformation proceeds as follows:

\begin{enumerate}[(1)]
\item Factorize the input polynomials $G$ into a set of pairwise coprime factors (with rational coefficients) $Q = \{q_1, \dots, q_n\}$. This also removes redundant/duplicate polynomials.

Each input polynomial $g \in G$ can be expressed in the form $g = c \prod_{i=1}^n q_i^{d_i}$ for some rational coefficient $c$ and natural number exponents $d_i \geq 1$ so the sign of $g$ is directly recovered from the signs of the factors $q\in Q$.
For example, if $g_1 = q_1q_2$ and in a consistent sign assignment $q_1$ is positive while $q_2$ is negative, then $g_1$ is negative according to that assignment, and so on.
Accordingly, to determine the set of all consistent sign assignments for $G$ it suffices to determine the same for $Q$.

\item Because the $q_i$'s are pairwise coprime, there is no consistent sign assignment where two or more $q_i$'s are set to zero.
So, in any given sign assignment, there is either \textit{exactly one} $q_i$ set to zero, or the $q_i$'s are \textit{all assigned to nonzero} (i.e., +1, -1) signs.

Now, for each $1 \leq i \leq n$, solve the restricted sign determination problem for all consistent sign assignments of $\{q_1, \dots, q_n\} \setminus \{q_i\}$ at the roots of $q_i$.
This yields all consistent sign assignments of $Q$ where exactly one $q_i$ is assigned to zero.

\item This step and the next step focus on finding all consistent sign assignments where all $q_i$'s are nonzero.
Compute a polynomial $p$ that satisfies the following properties:
\begin{enumerate}[\it i)]
\item $p$ is pairwise coprime with all of the $q_i$'s,
\item $p$ has a root in every interval between any two roots of the $q_i$'s,
\item $p$ has a root that is greater than all of the roots of the $q_i$'s, and
\item $p$ has a root that is smaller than all of the roots of the $q_i$'s.
\end{enumerate}

An explicit choice of $p$ satisfying these properties when $q_i \in Q$ are squarefree and pairwise coprime is shown in \rref{sec:formalize_preprocessing}.
The relationship between the roots of $p$ and the roots of $q_i \in Q$ is visualized in \rref{fig:rootsvis}.
Intuitively, the roots of $p$ (red points) provide representative sample points between the roots of the $q_i$'s (black squares).
\begin{figure}[hbt!]
\centering
\includegraphics[scale=0.3]{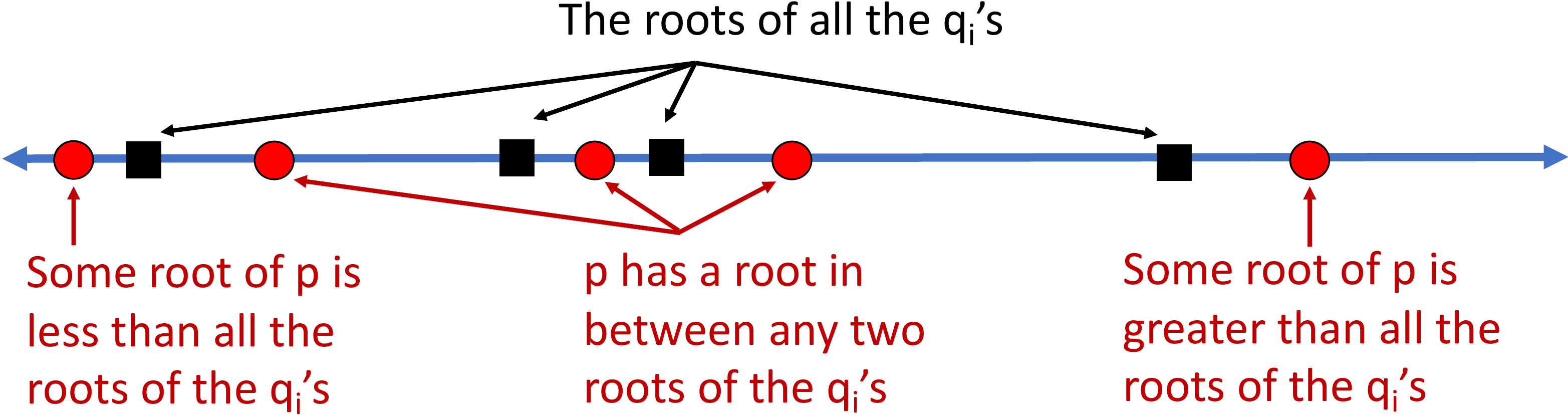}
\caption{The relation between the roots of the added polynomial $p$ and the roots of the $q_i$'s.}
\label{fig:rootsvis}
\end{figure}

\item Solve the restricted sign determination problem for all consistent sign assignments of $\{q_1, \dots, q_n\}$ at the roots of $p$.

Returning to~\rref{fig:rootsvis}, the $q_i$'s are sign-invariant in the intervals between any two roots of the $q_i$'s (black squares) and to the left and right beyond all roots of the $q_i$'s.
Intuitively, this is true because moving along the blue real number line in~\rref{fig:rootsvis}, no $q_i$ can change sign without first passing through a black square.
Thus, all consistent sign assignments of $q_i$ that only have nonzero signs must occur in one of these intervals and therefore, by sign-invariance, also at one of the roots of $p$ (red points).

\item The combined set of sign assignments where some $q_i$ is zero, as found in (2), and where no $q_i$ is zero, as found in (4), solves the sign determination problem for $Q$, and therefore also for $G$, as argued in (1).
\end{enumerate}

Our algorithm to solve the restricted sign determination problem using BKR's key insight is called $\isa{find\isacharunderscore consistent\isacharunderscore signs\isacharunderscore at\isacharunderscore roots}$; we now turn to the details of this method.

\subsection{Restricted Sign Determination}\label{sec:keystep}
The restricted sign determination problem for polynomials $q_1, \dots, q_n$ at the roots of a polynomial $p \neq 0$, where each $q_1,\dots,q_n$ is coprime with $p$, can be tackled naively by setting up and solving a \emph{matrix equation}.
The idea of using a matrix equation for sign determination dates back to Tarski \cite{Tarski} \cite[Section 10.3]{algRAG}, and accordingly our formalization shares some similarity to Cohen and Mahboubi's formalization \cite{AssiaQE} of Tarski's algorithm (see \cite[Section 11.2]{cohen_phd}).
BKR's additional insight is to avoid the prohibitive complexity of enumerating exponentially many possible sign assignments for $q_1, \dots, q_n$ by computing the matrix equation recursively and performing a \emph{reduction} that retains only the consistent sign assignments at each recursive step.
This reduction keeps intermediate data sizes manageable because the number of consistent sign assignments is bounded by the number of roots of $p$ throughout.
We first explain the technical underpinnings of the matrix equation before returning to our implementation of BKR's recursive procedure.
 \emph{For brevity, references to sign assignments for $q_1, \dots, q_n$ in this section are always at the roots of $p$}.

\subsubsection{Matrix Equation}\label{sec:mtxeqn}
The inputs to the matrix equation are a set of candidate (i.e., not necessarily consistent) sign assignments $\tilde{\Sigma} = \{ \tilde{\sigma}_1, \dots, \tilde{\sigma}_m \}$ for the polynomials $q_1, \dots, q_n$ and a set of subsets $S = \{I_1, \dots, I_l\}$, $I_i \subseteq \{1,\dots, n \}$ of indices selecting among those polynomials.
The set of all consistent sign assignments $\Sigma$ for $q_1, \dots, q_n$ is assumed to be a subset of $\tilde{\Sigma}$, i.e., $\Sigma \subseteq \tilde{\Sigma}$.

For example, consider $p = x^3 - x$ and $q_1 = 3x^3+2$.
The set of all possible candidate sign assignments $\tilde{\Sigma} = \{(+1), (-1)\}$ must contain the consistent sign assignments for $q_1$ (sign $(0)$ is impossible as $p, q_1$ are coprime).
The possible subsets of indices are $I_1 = \{\}$ and $I_2 = \{ 1 \}$.

The main algebraic tool underlying the matrix equation is the \emph{Tarski query} which provides semantic information about the number of roots of $p$ with respect to another polynomial $q$.
\begin{definition}
Given univariate polynomials $p,q$ with $p\neq0$, the \emph{Tarski query} $N(p, q)$ is:
\begin{align*}
    N(p, q) =&\ \#\{ x \in \mathbb{R} ~|~ p(x) =0, q(x) > 0 \} - \#\{ x \in \mathbb{R} ~|~ p(x) =0, q(x) < 0\}.
\end{align*}
\end{definition}

Importantly, the Tarski query $N(p,q)$ can be computed from input polynomials $p, q$ using Euclidean remainder sequences \emph{without} explicitly finding the roots of $p$.
This is a consequence of the Sturm-Tarski theorem which has been formalized in \Isabelle by Li~\cite{Sturm_Tarski-AFP}.
The theoretical complexity for computing $N(p, q)$ is $O(\deg p~(\deg p + \deg q))$ \cite[Sections 2.2.2 and 8.3]{algRAG}.
However, this complexity analysis does not take into account the growth in bitsizes of coefficients in the remainder sequences \cite[Section 8.3]{algRAG}, so it will not be not achieved by the current \Isabelle formalization of Tarski queries~\cite{Sturm_Tarski-AFP} without further optimization.

For the matrix equation, we lift Tarski queries to a \textit{subset} of the input polynomials:
\begin{definition}
Given a univariate polynomial $p \neq 0$, univariate polynomials $q_1, \dots, q_n$, and a subset $I \subseteq \{1, \dots, n\}$, the \emph{Tarski query} $N(I)$ with respect to $p$ is:
\begin{align*}
    N(I) = N(p,  \Pi_{i \in I} 	q_i) = &\ \#\{ x \in \mathbb{R} ~|~ p(x) =0, \Pi_{i \in I} q_i(x) > 0 \} \\ &- \#\{ x \in \mathbb{R} ~|~ p(x) =0, \Pi_{i \in I} 	q_i(x) < 0 \}.
\end{align*}
\end{definition}

The \emph{matrix equation} is the relationship $M \cdot w = v$ between the following three entities:
\begin{itemize}
\item $M$, the $l$-by-$m$ matrix with entries $M_{i,j} = \Pi_{k \in I_i} \tilde{\sigma}_j(q_k)\in\{-1,1\}$ for $I_i \in S$ and $\tilde{\sigma}_j \in \tilde{\Sigma}$,
\item $w$, the length $m$ vector whose entries count the number of roots of $p$ where $q_1, \dots, q_n$ has sign assignment $\tilde{\sigma}$, i.e., $ w_i = \#\{x \in \mathbb{R} ~|~ p(x) = 0, \text{sgn}(q_j(x)) = \tilde{\sigma}_i(q_j) ~\text{for all}~ 1\leq j \leq n\}$,
\item $v$, the length $l$ vector consisting of Tarski queries for the subsets, i.e., $v_i =  N(I_i)$.
\end{itemize}

Observe that the vector $w$ is such that the sign assignment $\tilde{\sigma}_i$ is consistent (at a root of $p$) iff its corresponding entry $w_i$ is nonzero.
Thus, the matrix equation can be used to solve the sign determination problem by solving for $w$.
In particular, the matrix $M$ and the vector $v$ are both computable from the input (candidate) sign assignments and subsets.
Further, since the subsets will be chosen such that the constructed matrix $M$ is \emph{invertible}, the matrix equation uniquely determines $w$ and the nonzero entries of $w = M^{-1} \cdot v$.

The following \Isabelle theorem summarizes sufficient conditions on the list of sign assignments \isa{signs} and the list of index subsets \isa{subsets} for the matrix equation to hold for polynomial list \isa{qs} at the roots of polynomial \isa{p}.
Note the switch from set-based representation to list-based representation in the theorem.
This formally provides an ordering to the polynomials, sign assignments, and subsets, which is useful for computations.
\begin{isabelle}
\matrixequationthm
\end{isabelle}

Here, \isa{M\isacharunderscore mat}, \isa{w\isacharunderscore vec}, and \isa{v\isacharunderscore vec} construct the matrix $M$ and vectors $w$, $v$ respectively; \isa{*$_v$} denotes the matrix-vector product in \Isabelle.
The switch into list notation necessitates some consistency assumptions, e.g., that the \isa{signs} list contains \isa{distinct} sign assignments and that the index \isa{i} occurring in each list of indices \isa{l} in \isa{subsets} points to a valid element of the list \isa{qs}.
The proof of \isa{matrix\isacharunderscore equation} uses a counting argument: intuitively, $M_{i,j}$ is the contribution of any real value $x$ that has the sign assignment $\tilde{\sigma}_j$ towards $N(I_i)$, so multiplying these contributions by the actual counts of those real values in $w$ gives $M_i \cdot w = v_i$.

Note that the theorem does \emph{not} ensure that the constructed matrix $M$ is invertible (or even square).
This must be ensured separately when solving the matrix equation for $w$.
We now discuss BKR's inductive construction and its usage of the matrix equation.

\subsubsection{Base Case}\label{sec:basecase}
\begin{figure}[t]
\centering
\includegraphics[scale=0.45]{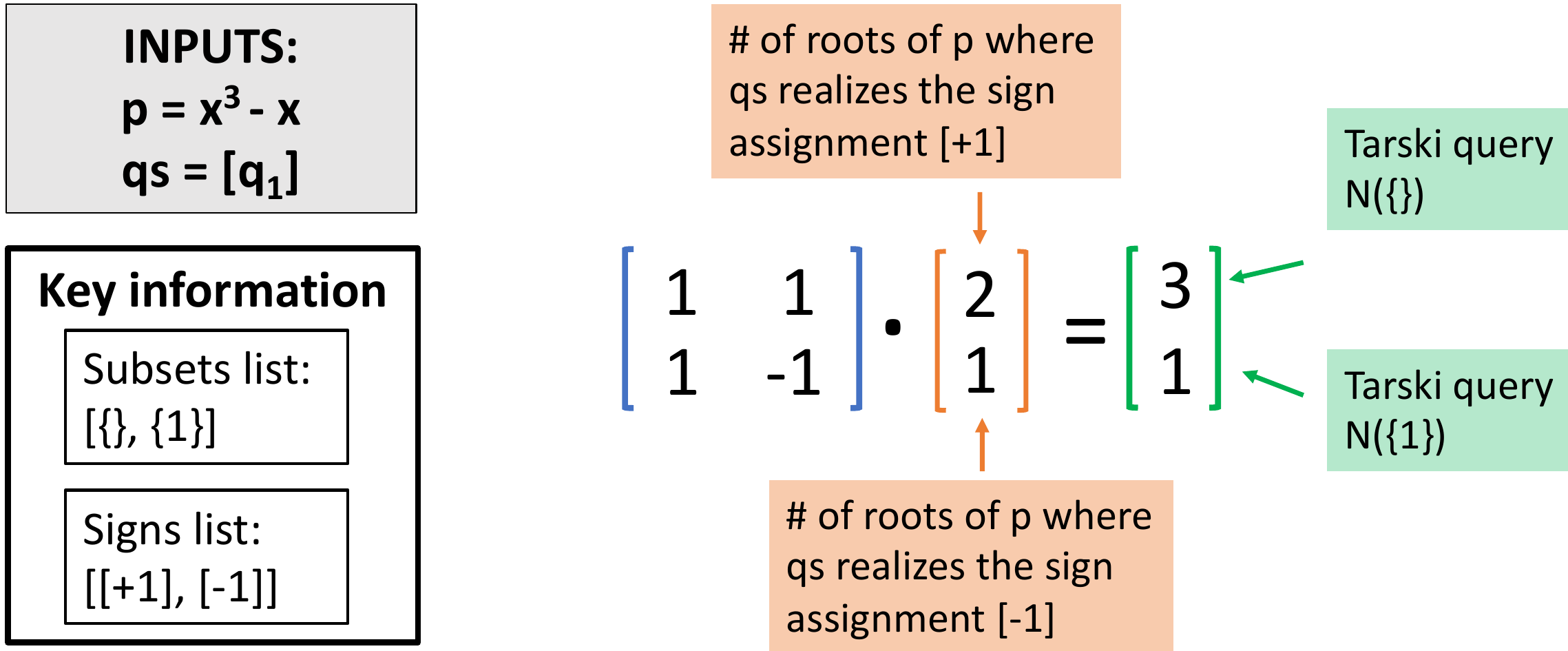}
\caption{Matrix equation for $p = x^3 - x$, $q_1 = 3x^3+2$.} 
\label{fig:matrixeq}
\end{figure}

The simplest (base) case of the algorithm is when there is a single polynomial $[q_1]$.
Here, it suffices to set up a matrix equation $M\cdot w = v$ from which we can compute all consistent sign assignments.
As hinted at earlier, this can be done with the list of index subsets $[\{\},\{1\}]$ and the candidate sign assignment list $[(+1), (-1)]$.\footnote{In the \Isabelle formalization, we use $0$-indexed lists to represent sets and sign assignments, so the subsets list is represented as \isa{[[],[0]]} and the signs list is \isa{[[1],[-1]]}.}
Further, as illustrated in~\rref{fig:matrixeq}, the matrix $M$ is invertible for these choices of subsets and candidate sign assignments, so the matrix equation can be explicitly solved for $w$.

\subsubsection{Inductive Case: Combination Step}\label{sec:combining}

The matrix equation can be similarly used to determine the consistent sign assignments for an arbitrary list of polynomials $[q_1,\dots,q_n]$.
The driving idea for BKR is that, given two solutions of the sign determination problem at the roots of $p$ for two input lists of polynomials, say, $\ell_1 = [ r_1, \dots, r_k ]$ and $\ell_2 = [ r_{k+1},\dots,r_{k+l} ]$, one can combine them to yield a solution for the list of polynomials $[r_1,\dots,r_{k+l}]$.
This yields a recursive method for solving the sign determination problem by solving the base case at the single polynomials $[q_1], [q_2], \dots, [q_n]$, and then recursively combining those solutions, i.e., solving $[q_1,q_2], [q_3,q_4], \dots$, then $[q_1,q_2,q_3,q_4], \dots$, and so on until a solution for $[q_1,\dots,q_n]$ is obtained.
Importantly, BKR performs a reduction (\rref{sec:reduction}) after each combination step to bound the size of the intermediate data.

More precisely, assume for $\ell_1$, we have a list of index subsets $S_1$ and a list of sign assignments $\tilde{\Sigma}_1$ such that $\tilde{\Sigma}_1$ contains all of the consistent sign assignments for $\ell_1$ and the matrix $M_1$ constructed from $S_1$ and $\tilde{\Sigma}_1$ is invertible.
Accordingly, for $\ell_2$, we have the list of subsets $S_2$, list of sign assignments $\tilde{\Sigma}_2$ containing all consistent sign assignments for $\ell_2$, and $M_2$ constructed from $S_2$, $\tilde{\Sigma}_2$ is invertible.
In essence, we are assuming that $S_1, \tilde{\Sigma}_1$ and $S_2, \tilde{\Sigma}_2$ satisfy the hypotheses for the matrix equation to hold, so that they contain all the information needed to solve for the consistent sign assignments of $\ell_1$ and $\ell_2$ respectively.

Observe that any consistent sign assignment for $\ell \mnodefeq [r_1,\dots,r_{k+l}]$ must have a prefix that is itself a consistent sign assignment to $\ell_1$ and a suffix that is itself a consistent sign assignment to $\ell_2$.
Thus, the combined list of sign assignments $\tilde{\Sigma}$ obtained by concatenating every entry of $\tilde{\Sigma}_1$ with every entry of $\tilde{\Sigma}_2$ necessarily contains all consistent sign assignments for $\ell$.
The combined subsets list $S$ is obtained in an analogous way from $S_1$, $S_2$ (where concatenation is now set union), with a slight modification: the subset list $S_2$ indexes polynomials from $\ell_2$, but those polynomials now have different indices in $\ell$, so everything in $S_2$ is shifted by the length of $\ell_1$ before combination.
Once we have the combined subsets list, we can calculate the RHS vector $v$ with Tarski queries as explained in \rref{sec:mtxeqn}.
%

\begin{figure}[t]
\centering
\includegraphics[scale=0.35]{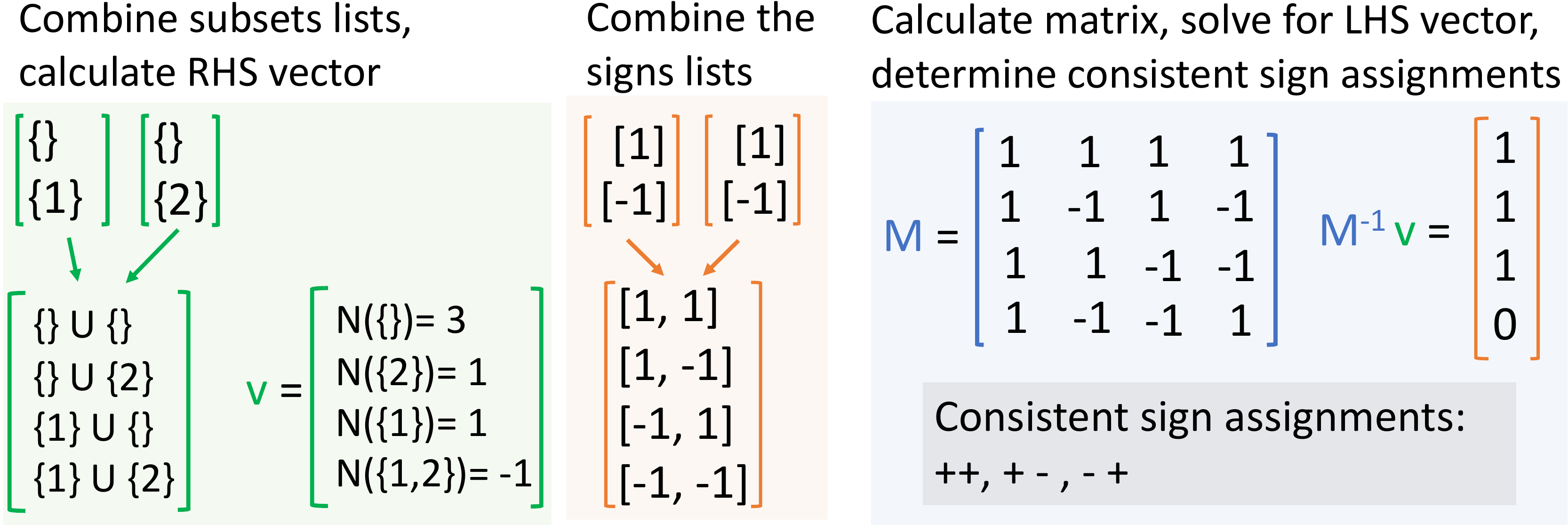}
\caption{Combining two systems.}
\label{fig:smash}
\end{figure}

The matrix $M$ constructed from $S$, $\tilde{\Sigma}$ is exactly the Kronecker product of $M_1$ and $M_2$.
Further, the Kronecker product of invertible matrices is invertible, so the matrix equation can be solved for the LHS vector $w$ using $M$ and the vector $v$ computed from the subsets list $S$.
Then the nonzero entries of $w$ correspond to the consistent sign assignments of $\ell$.
Taking a concrete example, suppose we want to find the list of consistent sign assignments for $\ell = [3x^3 + 2, 2x^2 - 1]$ at the zeros of $p = x^3 - x$.
The combination step for $\ell_1 = [3x^3 + 2]$ and $\ell_2 = [2x^2 - 1]$ is visualized in \rref{fig:smash}.


\subsubsection{Reduction Step}\label{sec:reduction}
The reduction step takes an input list of index subsets $S$ and candidate sign assignments $\tilde{\Sigma}$.
It removes the inconsistent sign assignments and then unnecessary index subsets, which keeps the size of the intermediate data tracked for the matrix equation as small as possible.

The reduction step is best explained in terms of the matrix equation $M \cdot w = v$ constructed from the inputs $S$, $\tilde{\Sigma}$.
After solving for $w$, the reduction starts by deleting all indexes of $w_i$ that are $0$ and the corresponding $i$-th sign assignments in $\tilde{\Sigma}$ which are now known to be inconsistent (recall that $w_i$ counts the number of zeros of $p$ where the $i$-th sign assignment is realized).
This corresponds to deleting the $i$-th columns of matrix $M$.
If any columns are deleted, the resulting matrix is no longer square (nor invertible).
Thus, the next step finds a basis among the remaining rows of the matrix to make it invertible again (deleting any rows that do not belong to the chosen basis).
Deleting the $j$-th row in this matrix corresponds to deleting the $j$-th index subset in $S$.

\begin{figure}[t]
\centering
\includegraphics[scale=0.35]{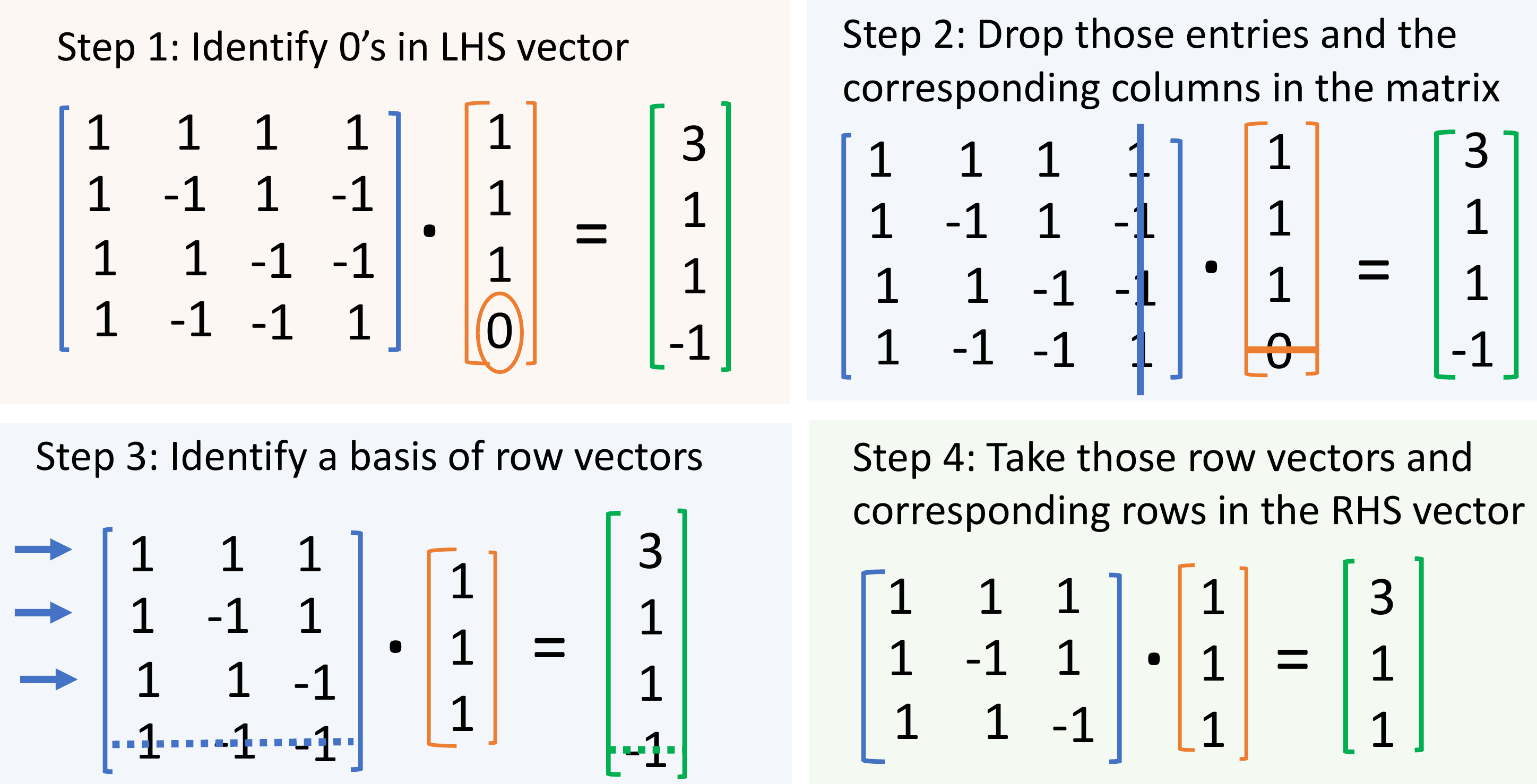}
\caption{Reducing a system.}
\label{fig:reduction}
\end{figure}

The reduction step for the matrix equation with $p = x^3 - x$ and $\ell = [3x^3 + 2, 2x^2 - 1]$ is visualized in \rref{fig:reduction}.
Naively using the matrix equation for restricted sign determination would require $2^{|\ell|} = 4$ Tarski queries for this example, whereas $2+2+4 = 8$ queries are required using BKR ($2$ for each base case, $4$ for the combination step).
However, for longer lists $\ell$, the naive approach requires $2^{|\ell|}$ queries while BKR's reduction step ensures that the number of intermediate consistent sign assignments is bounded by the number of roots of $p$ (and hence $\deg p$) throughout.
This difference is shown in~\rref{sec:code} and is also illustrated by~\rref{fig:reduction}, where $p$ has degree $3$ and there are $3$ consistent sign assignments for $\ell$ after reduction.

\section{Formalization}\label{sec:formalization}
Now that we have set up the theory behind the BKR algorithm, we turn to some details of our formalization: the proofs, extensions to the existing matrix libraries, and the exported code.
Our proof builds significantly on existing proof developments in the Archive of Formal Proofs~\cite{Sturm_Tarski-AFP,Jordan_Normal_Form-AFP,Algebraic_Numbers-AFP}.
\Isabelle's builtin search tool and Sledgehammer~\cite{DBLP:conf/lpar/PaulsonB10} provided invaluable automation for discovering existing theorems and for finishing (easy) subgoals in proofs.
The most challenging part of the formalization, in our opinion, is the reduction step, in no small part because it involves significant linear algebra (further details in \rref{sec:matrixreduction}).

\subsection{Formalizing the Decision Procedure}
In this section, we discuss the proofs for our decision procedure in \emph{reverse} order compared to \rref{sec:mathematics}; that is, we first discuss the formalization of our algorithm for restricted sign determination \isa{find\isacharunderscore consistent\isacharunderscore signs\isacharunderscore at\isacharunderscore roots} before discussing the top-level decision procedures for univariate real arithmetic, \isa{decide\isacharunderscore \{universal|existential\}}.
The reader may wish to revisit~\rref{sec:mathematics} for informal intuition behind the procedure while reading this section. 

\subsubsection{Sign Determination at Roots}\label{sec:signdetAtRoots}
We combine BKR's base case (\rref{sec:basecase}), combination step (\rref{sec:combining}), and reduction step (\rref{sec:reduction}) to form our core algorithm \isa{calc\isacharunderscore data} for the restricted sign determination problem at the roots of a polynomial.
The \isa{calc\isacharunderscore data} algorithm takes a real polynomial \isa{p} and a list of polynomials \isa{qs} and produces a $3$-tuple \isa{(M, S, $\Sigma$)}, consisting of the matrix \isa{M} from the matrix equation, the list of index subsets \isa{S}, and the list of all consistent sign assignments \isa{$\Sigma$} for \isa{qs} at the roots of \isa{p}.
Although \isa{M} can be calculated directly from \isa{S} and \isa{$\Sigma$}, it is returned (as part of the algorithm), to avoid redundantly recomputing it at every recursive call.
\begin{isabelle}
\calcdatadef
\end{isabelle}
\begin{isabelle}
\findconsistentsignsatrootsdef
\end{isabelle}

The base case where \isa{qs} has length $\leq 1$ is handled\footnote{The trivial case where \isa{length qs = 0} is also handled for completeness; in this case, the list of consistent sign assignments is empty if $p$ has no real roots, otherwise, it is the singleton list \isa{[[]]}.} using the fixed choice of matrix, index subsets, and sign assignments (defined as the constant \isa{base\isacharunderscore case\isacharunderscore info}) from~\rref{sec:basecase}.
Otherwise, when \isa{length qs $> 1$}, the list is partitioned into two sublists \isa{qs1, qs2} and the algorithm recurses on those sublists.
The outputs for both sublists are combined using \isa{combine\isacharunderscore systems} which takes the Kronecker product of the output matrices and concatenates the index subsets and sign assignments as explained in~\rref{sec:combining}.
Finally, \isa{reduce\isacharunderscore system} performs the reduction according to~\rref{sec:reduction}, removing inconsistent sign assignments and redundant subsets of indices.
The top-level procedure is \isa{find\isacharunderscore consistent\isacharunderscore signs\isacharunderscore at\isacharunderscore roots}, which returns only \isa{$\Sigma$} (the third component of \isa{calc\isacharunderscore data}).
The following \Isabelle snippets show its main correctness theorem and important relevant definitions.
\begin{isabelle}
\rootsdef
\end{isabelle}

\begin{isabelle}
\consistentsignsatrootsdef
\end{isabelle}

\begin{isabelle}
\findconsistentsignsatrootsthm
\end{isabelle}

Here, \isa{roots} defines the set of roots of a polynomial \isa{p} (non-constructively), i.e., real values \isa{x} where the polynomial evaluates to 0 (\isa{poly p x = 0}).
Similarly, \isa{consistent\isacharunderscore signs\isacharunderscore at\isacharunderscore roots} returns the set of all sign vectors for the list of polynomials \isa{qs} at the roots of \isa{p}; \isa{sgn\isacharunderscore vec} returns the sign vector for input \isa{qs} at a real value and \isa{\isacharbackquote} is \Isabelle notation for the image of a function on a set.
These definitions are not meant to be computational.
Rather, they are used to state the correctness theorem that the algorithm \isa{find\isacharunderscore consistent\isacharunderscore signs\isacharunderscore at\isacharunderscore roots} (and hence \isa{calc\isacharunderscore data}) computes \emph{exactly all} consistent sign assignments for \isa{p} and \isa{qs} for input polynomial \isa{p $\not=$ 0} and polynomial list \isa{qs}, where every entry in \isa{qs} is \isa{coprime} to \isa{p}.

The proof of \isa{find\isacharunderscore consistent\isacharunderscore signs\isacharunderscore at\isacharunderscore roots} is by induction on \isa{calc\isacharunderscore data}.
Specifically, we prove that the following properties (our inductive invariant) are \textit{satisfied by the base case and maintained by both the combination step and the reduction step}:

\begin{enumerate}
\item The signs list is well-defined, i.e., the length of every entry in the signs list is the same as the length of the corresponding \isa{qs}. Additionally, all assumptions on \isa{S} and \isa{$\Sigma$} from the \isa{matrix\isacharunderscore equation} theorem from \rref{sec:mtxeqn} hold. (In particular, the algorithm always maintains a distinct list of sign assignments that, when viewed as a set, is a superset of all consistent sign assignments for \isa{qs}.)
\item The matrix \isa{M} matches the matrix calculated from \isa{S} and \isa{$\Sigma$}.
(Since we do not directly compute the matrix from \isa{S} and \isa{$\Sigma$}, as defined in~\rref{sec:mtxeqn}, we need to verify that our computations keep track of \isa{M} correctly.)
\item The matrix \isa{M} is invertible (so $M\cdot w = v$ can be uniquely solved for $w$).
\end{enumerate}

Some of these properties are easier to verify than others. The well-definedness properties, for example, are quite straightforward.
In contrast, matrix invertibility is more complicated to verify, especially after the reduction step; we will discuss this in more detail in \rref{sec:matrix}.
The inductive invariant establishes that we have a superset of the consistent sign assignments throughout the construction.
This is because the base case and the combination step may include extraneous sign assignments.
Only the reduction step is guaranteed to produce exactly the set of consistent sign assignments.
Thus the other main ingredient in our formalization, besides the inductive invariant, is a proof that the reduction step deletes all inconsistent sign assignments.
As \isa{calc\isacharunderscore data} always calls the reduction step before returning output, \isa{calc\isacharunderscore data} returns exactly the set of all consistent sign assignments, as desired.

\subsubsection{Building the Univariate Decision Procedure}\label{sec:formalize_preprocessing}
To prove the $\isa{decision\isacharunderscore procedure}$ theorem from \rref{sec:QEtoSD}, we need to establish correctness of \isa{find\isacharunderscore consistent\isacharunderscore signs}.
The most interesting part is formalizing the transformation described in \rref{sec:settingup}.
We discuss the steps from~\rref{sec:settingup} enumerated (1)--(5) below.

\begin{enumerate}[(1)]
\item Our procedure takes an input list of rational polynomials $G = [g_1, \dots, g_k]$ and computes a list of their pairwise coprime and squarefree factors\footnote{This is actually overkill: we do not necessarily need to \emph{completely} factor every polynomial in $G$ to transform $G$ into a set of pairwise coprime factors.
BKR suggest a parallel algorithm based in part on the literature \cite{10.1145/800061.808728} to find a ``basis set'' of squarefree and pairwise coprime polynomials.} $Q = [q_1, \dots, q_n]$.
An efficient method to factor a \emph{single} rational polynomial is formalized in \Isabelle by Divas\'{o}n et al.~\cite{10.1145/3018610.3018617}; we slightly modified their proof to find factors for a \emph{list} of polynomials while ensuring that the resulting factors are pairwise coprime, which implies that their product $\prod_i q_i$ is squarefree.

\item This step makes $n$ calls to \isa{find\isacharunderscore consistent\isacharunderscore signs\isacharunderscore at\isacharunderscore roots}, one for each $Q \setminus \{q_i\}$.

\item We choose the polynomial $p = (x - \isa{crb} (\prod_i q_i))(x+\isa{crb}  (\prod_i q_i)) (\prod_i q_i)'$, where $(\prod_i q_i)'$ is the formal polynomial derivative of $\prod_i q_i$ and $\isa{crb}(\prod_i q_i)$ is a computable positive integer with larger magnitude than any real root of $\prod_i q_i$.
The choice of $\isa{crb} (\prod_i q_i)$ uses a proof of the Cauchy root bound \cite[Section 10.1]{algRAG} by Thiemann and Yamada~\cite{Algebraic_Numbers-AFP}.
We prove that $p$ satisfies the four properties of step (3) from \rref{sec:settingup}:
\begin{enumerate}[\it i)]
\item Since $\prod_i q_i$ is squarefree, $(\prod_i q_i)'$ is coprime with $\prod_i q_i$ and, thus, also coprime with each of the $q_i$'s.
Because $\isa{crb} (\prod_i q_i)$ is strictly larger in magnitude than all of the roots of the roots of the $q_i$'s, it follows that $p$ is also coprime with all of the $q_i$'s.

\item By Rolle's theorem\footnote{For differentiable function $f:\mathbb{R}\mapsto \mathbb{R}$ with $f(a) = f(b)$, $a < b$, there exists $a < z < b$ where $f'(z) = 0$.} (which is already formalized in Isabelle/HOL's standard library), $(\prod_i q_i)'$ has a root between every two roots of $\prod_i q_i$ and therefore \isa{p} also has a root in every interval between any two roots of the $q_i$'s.

\item and {\it iv)} This choice of $p$ has roots at $-\isa{crb} (\prod_i q_i)$ and $\isa{crb} (\prod_i q_i)$, which are respectively smaller and greater than all roots of the $q_i$'s.
\end{enumerate}


\item Each polynomial $q_i$ is sign invariant between its roots.\footnote{By the intermediate value theorem (which is already formalized in Isabelle/HOL's standard library), if $q_i$ changes sign, e.g., from positive to negative, between two adjacent roots, then there exists a third root in between those adjacent roots, which is a contradiction.} Accordingly, the $q_i$'s are sign invariant between the roots of $\prod_i q_i$ (and to the left/right of all roots of the $q_i$'s). 

\item We use the \isa{find\isacharunderscore consistent\isacharunderscore signs\isacharunderscore at\isacharunderscore roots} algorithm with $Q$ and our chosen $p$.
\end{enumerate}

Putting the pieces together, we verify that \isa{find\isacharunderscore consistent\isacharunderscore signs} finds exactly the consistent sign assignments for its input polynomials.
The \isa{decision\isacharunderscore procedure} theorem follows by induction over the \isa{fml} type representing formulas of univariate real arithmetic and our formalized semantics for those formulas.

\subsection{Matrix Library}\label{sec:matrix}
Matrices feature prominently in our algorithm: the combination step uses the Kronecker product, while the reduction step requires matrix inversion and an algorithm for finding a basis from the rows (or, equivalently, columns) of a matrix.
There are a number of linear algebra libraries available in \Isabelle~\cite{Rank_Nullity_Theorem-AFP,Matrix-AFP,Jordan_Normal_Form-AFP}, each building on a different underlying representation of matrices.
We use the formalization by Thiemann and Yamada~\cite{Jordan_Normal_Form-AFP} as it provides most of the matrix algorithms required by our decision procedure and supports efficient code extraction~\cite[Section 1]{Jordan_Normal_Form-AFP}.
Naturally, any such choice leads to tradeoffs; we now detail some challenges of working with the library and some new results we prove.

\subsubsection{Combination Step: Kronecker Product}
We define the Kronecker product for matrices \isa{A}, \isa{B} over a \isa{ring} as follows:
\begin{isabelle}
\kroneckerproductdef
\end{isabelle}

Matrices with entries of type \isa{'a} are constructed with \isa{mat m n f}, where \isa{m, n :: nat} are the number of rows and columns of the matrix respectively, and \isa{f :: nat $\times$ nat $\Rightarrow$ 'a} is such that \isa{f i j} gives the matrix entry at position \isa{i}, \isa{j}.
Accordingly, \isa{M \$\$ (i,j)} extracts the \isa{(i,j)}-th entry of matrix \isa{M}, and \isa{dim\isacharunderscore row, dim\isacharunderscore col} return the number of rows and columns of a matrix respectively.

We prove basic properties of our definition of the Kronecker product: it is associative, distributes over addition, and satisfies the mixed-product identity for matrices \isa{A}, \isa{B}, \isa{C}, \isa{D} with compatible dimensions (for \isa{A * C} and \isa{B * D}): 
\begin{align*}
\isa{kronecker\isacharunderscore product}& \isa{ (A * C) (B * D) =} \\ &\isa{(kronecker\isacharunderscore product A B) * (kronecker\isacharunderscore product C D)}.
\end{align*}
The mixed-product identity implies that the Kronecker product of invertible matrices is invertible.
Briefly, for invertible matrices \isa{A}, \isa{B} with respective inverses \isa{A$^{-1}$}, \isa{B$^{-1}$}, the mixed product identity gives:
\begin{align*}
\isa{(kronecker\isacharunderscore product A B) *}& \isa{ (kronecker\isacharunderscore product A$^{-1}$ B$^{-1}$) =} \\
&\isa{kronecker\isacharunderscore product (A * A$^{-1}$) (B *  B$^{-1}$) = I},
\end{align*}
 where \isa{I} is the identity matrix.
In other words, \isa{kronecker\isacharunderscore product A B} and \isa{kronecker\isacharunderscore product A$^{-1}$ B$^{-1}$} are inverses.
We use this to prove that the matrix obtained by the combination step is invertible (part of the inductive hypothesis from \rref{sec:signdetAtRoots}).

\begin{remark}
Prathamesh~\cite{Matrix_Tensor-AFP} formalized Kronecker products for \Isabelle's default matrix type. For computational purposes, we provide a new formalization that is compatible with the matrix representation of Thiemann and Yamada~\cite{Jordan_Normal_Form-AFP}.
\end{remark}

\subsubsection{Reduction Step: Gauss--Jordan and Matrix Rank}\label{sec:matrixreduction}
Our reduction step makes extensive use of the Gauss--Jordan elimination algorithm by Thiemann and Yamada~\cite{10.1145/2854065.2854073}.
First, we use matrix inversion based on Gauss--Jordan elimination to invert the matrix $M$ in the matrix equation (\rref{sec:mtxeqn} and Step 1 in~\rref{fig:reduction}).
We also contribute new proofs surrounding their Gauss--Jordan elimination algorithm in order to use it to extract a basis from the rows (equivalently columns) of a matrix (Step 3 in~\rref{fig:reduction}).

Suppose that an input matrix \isa{A} has more rows than columns, e.g., the matrix in Step 2 of~\rref{fig:reduction}.
The following definition of \isa{rows\isacharunderscore to\isacharunderscore keep} returns a list of (distinct) row indices of \isa{A}.
\begin{isabelle}
\rowstokeepdef
\end{isabelle}

Here, \isa{gauss\isacharunderscore jordan\isacharunderscore single} returns the row-reduced echelon form (RREF) of \isa{A} after Gauss--Jordan elimination and \isa{pivot\isacharunderscore positions} finds the positions, i.e., \isa{(row, col)} pairs, of the first nonzero entry in each row of the matrix; both are existing definitions from the library by Thiemann and Yamada~\cite{10.1145/2854065.2854073}.
Our main new result for \isa{rows\isacharunderscore to\isacharunderscore keep} is:
\begin{isabelle}
\rowstokeepranklem
\end{isabelle}

Here \isa{vec\isacharunderscore space.rank n M} (defined by Bentkamp~\cite{Jordan_Normal_Form-AFP}) is the finite dimension of the vector space spanned by the columns of \isa{M}.
Thus, the lemma says that keeping only the pivot rows of matrix \isa{A} (with \isa{take\isacharunderscore rows A (rows\isacharunderscore to\isacharunderscore keep A)}) preserves the rank of \isa{A}.
At a high level, the proof of \isa{rows\isacharunderscore to\isacharunderscore keep\isacharunderscore rank} is in three steps:
\begin{enumerate}
\item First, we prove a version of \isa{rows\isacharunderscore to\isacharunderscore keep\isacharunderscore rank} for the pivot \emph{columns} of a matrix and where \isa{A} is assumed to be a matrix in RREF.
The RREF assumption for \isa{A} enables direct analysis of the shape of its pivot columns.
\item Next, we lift the result to an arbitrary matrix \isa{A}, which can always be put into RREF form by \isa{gauss\isacharunderscore jordan\isacharunderscore single}.
\item Finally, we formalize the following classical result that column rank is equal to row rank:
\isa{vec\isacharunderscore space.rank (dim\isacharunderscore row A) A = vec\isacharunderscore space.rank (dim\isacharunderscore col A) (A$^T$)}.
We lift the preceding results for pivot columns to also work for pivot rows by matrix transposition (pivot rows of matrix \isa{A} are the pivot columns of the transpose matrix \isa{A$^T$}).
\end{enumerate}

To complete the proof of the reduction step, recall that the matrix in Step 2 of~\rref{fig:reduction} is obtained by dropping columns of an invertible matrix.
The resulting matrix has full column rank but more rows than columns.
We show that when \isa{A} in \isa{rows\isacharunderscore to\isacharunderscore keep\isacharunderscore rank} has full column rank (its \isa{rank} is \isa{dim\isacharunderscore col A}) then \isa{length (rows\isacharunderscore to\isacharunderscore keep A) = dim\isacharunderscore col A} and so the matrix consisting of pivot rows of \isa{A} is square, has full rank, and is therefore invertible.

\begin{remark}
Divas\'{o}n and Aransay formalized the equivalence of row and column rank for \Isabelle's default matrix type~\cite{Gauss_Jordan-AFP} while we have formalized the same result for Bentkamp's definition of matrix rank~\cite{Jordan_Normal_Form-AFP}.
Another technical drawback of our choice of libraries is the locale argument \isa{n} for \isa{vec\isacharunderscore space}.
Intuitively (for real matrices) this carves out subsets of $\reals^\isa{n}$ to form the vector space spanned by the columns of \isa{M}.
Whereas one would usually work with \isa{n} fixed and implicit within an \Isabelle locale, we pass the argument explicitly here because our theorems often need to relate the rank of vector spaces in $\reals^\isa{m}$ and $\reals^\isa{n}$ for $\isa{m} \not= \isa{n}$.
This negates some of the automation benefits of \Isabelle's locale system.
\end{remark}

\subsection{Code Export}\label{sec:code}
We export our decision procedure to Standard ML, compile with \texttt{mlton}, and test it on 10 microbenchmarks from~\cite[Section 8]{li2019deciding}.
While we leave extensive experiments for future work since our implementation is unoptimized, we compare the performance of our procedure using BKR sign determination (Sections~\ref{sec:basecase}--\ref{sec:reduction}) versus an \textit{unverified} implementation that naively uses the matrix equation (\rref{sec:mtxeqn}).
We also ran Li \emph{et al.}'s \texttt{univ\isacharunderscore rcf} decision procedure~\cite{li2019deciding} which can be directly executed as a proof tactic in \Isabelle (code kindly provided by Wenda Li).
The benchmarks were ran on an Ubuntu 18.04 laptop with 16GB RAM and 2.70 GHz Intel Core i7-6820HQ CPU.
Results are in~\rref{tab:experiments}.

The most significant bottleneck in our current implementation is the computation of Tarski queries $N(p,q)$ when solving the matrix equation.
Recall for our algorithm (\rref{sec:mtxeqn}) the input $q$ to $N(p,q)$ is a \emph{product} of (subsets of) polynomials appearing in the inputs.
Indeed, \rref{tab:experiments} shows that the algorithm performs well when the factors have low degrees, e.g., ex1, ex2, ex4, and ex5. Conversely, it performs poorly on problems with many factors and higher degrees, e.g., ex3, ex6, and ex7.
Further, as noted in experiments by Li and Paulson~\cite{10.1145/3293880.3294092}, the Sturm-Tarski theorem in \Isabelle currently uses a straightforward method for computing remainder sequences which can also lead to significant (exponential) blowup in the bitsizes of rational coefficients of the involved polynomials.
This is especially apparent for ex6 and ex7, which have large polynomial degrees and high coefficient complexity; these time out without completing even a single Tarski query.
From~\rref{tab:experiments}, the BKR approach successfully reduces the number of Tarski queries as the number of input factors grows---the number of queries for BKR is dependent on the polynomial degrees and the number of consistent sign assignments, while the naive approach always requires exactly $(\frac{n}{2}+1)2^n$ queries for $n$ factors\footnote{For $n$ factors,~\rref{sec:settingup}'s transformation yields $n$ restricted sign determination subproblems involving $n-1$ polynomials each and one subproblem involving $n$ polynomials. Using naive sign determination to solve all of these subproblems requires $n(2^{n-1}) + 2^n = (\frac{n}{2}+1)2^n$ Tarski queries in total.} (which are reported in \rref{tab:experiments} whether completed or not).
On the other hand, there is some overhead for smaller problems, e.g., ex1, ex3, that arises from the recursion in BKR.

The \texttt{univ\isacharunderscore rcf} tactic relies on an external solver (we used Mathematica 12.1.1) to produce \emph{untrusted} certificates which are then formally checked (by reflection) in \Isabelle~\cite{li2019deciding}.
This procedure is optimized and efficient: except for ex7 where the tactic timed out, most of the time (roughly 3 seconds per example) is actually spent to start an instance of the external solver.

An important future step, e.g., to enable use of our procedure as a tactic in \Isabelle, is to avoid coefficient growth by using pseudo-division~\cite[Section 3]{NASATarski} or more advanced techniques: for example, using subresultants to compute polynomial GCDs (and thereby build the remainder sequences)~\cite{ducos2000optimizations}.
Pseudo-division is also important in the multivariate generalization of BKR (discussed in Appendix \ref{sec:multivar}), where the polynomial coefficients of concern are themselves (multivariate) polynomials rather than rational numbers.
The pseudo-division method has been formalized in \Isabelle~\cite{li2019deciding}, but it is not yet available on the AFP.

\begin{table}[t]
\centering
\begin{tabular}{lr@{~}rr@{~}rrrrrrrr}
\toprule
\textbf{Formula} & \multicolumn{2}{r}{\textbf{{\#}Poly}} & \multicolumn{2}{r}{\textbf{{\#}Factor}} & \textbf{\begin{tabular}[c]{@{}r@{}}\textbf{{\#}$\bm{N(p,q)}$}\\ (Naive)\end{tabular}} & \textbf{\begin{tabular}[c]{@{}r@{}}\textbf{{\#}$\bm{N(p,q)}$}\\ (BKR)\end{tabular}} & \textbf{\begin{tabular}[c]{@{}r@{}}Time\\ (Naive)\end{tabular}} & \textbf{\begin{tabular}[c]{@{}r@{}}Time\\ (BKR)\end{tabular}} & \textbf{\begin{tabular}[c]{@{}r@{}}Time\\ (\cite{li2019deciding})\end{tabular}}\\ \midrule
ex1                          & 4  &(12) & ~~ 3  & (1)  & 20           & 31  & 0.003      & 0.006     & 3.020 \\
ex2                          & 5  &(6)  & ~~ 7  & (1)  & 576          & 180 & 5.780      & 0.442     & 3.407 \\
ex3                          & 4  &(22) & ~~ 5  & (22) & 112          & 120 & 1794.843   & 1865.313  & 3.580 \\
ex4                          & 5  &(3)  & ~~ 5  & (2)  & 112          & 95  & 0.461      & 0.261     & 3.828 \\
ex5                          & 8  &(3)  & ~~ 7  & (3)  & 576          & 219 & 28.608     & 8.333     & 3.806 \\
ex6                          & 22 &(9)  & ~~ 22 & (8)  & 50331648     & -   & -          & -         & 6.187 \\
ex7                          & 10 &(12) & ~~ 10 & (11) & 6144         & -   & -          & -         & - \\
ex1 $\land$ 2                       & 9  &(12) & ~~ 9  & (1)  & 2816         & 298 & 317.432    & 3.027     & 3.033 \\
ex1 $\land$ 2 $\land$ 4                    & 13 &(12) & ~~ 12 & (2)  & 28672        & 555 & -          & 51.347    & 3.848 \\
ex1 $\land$ 2 $\land$ 5                    & 16 &(12) & ~~ 14 & (3)  & 131072       & 826 & -          & 436.575   & 3.711 \\ \bottomrule
\end{tabular}
\caption{Comparison of decision procedures using naive and BKR sign determination and Li \emph{et al.}'s \texttt{univ\isacharunderscore rcf} tactic in \Isabelle~\cite{li2019deciding}.
All formulas are labeled following~\cite[Section 8]{li2019deciding}; formulas with $\land$ indicate conjunctions of the listed examples.
Columns: \textbf{{\#}Poly} counts the number of distinct polynomials appearing in the formula (maximum degree among polynomials in parentheses), \textbf{{\#}Factor} counts the number of distinct factors from (1) in~\rref{sec:settingup} (maximum degree among factors in parentheses), \textbf{{\#}$\bm{N(p,q)}$} counts the number of Tarski queries made by each approach, and \textbf{Time} reports time taken (seconds, 3 d.p.) for each decision procedure to run to completion.
Cells with - indicate a timeout after 1 hour.}
\label{tab:experiments}
\end{table}

\section{Related Work}\label{sec:relatedwork}
Our work fits into the larger body of formalized univariate decision procedures.
Most closely related are Li \emph{et al.}'s formalization of a CAD-based univariate QE procedure in \Isabelle \cite{li2019deciding} and the $\isa{tarski}$ univariate QE strategy formalized in PVS \cite{NASATarski}.
We discuss each in turn.

The univariate CAD algorithm underlying Li \emph{et al.}'s approach~\cite{li2019deciding} decomposes $\mathbb{R}$ into a set of sign-invariant regions, so that every polynomial of interest has constant sign within each region.
A real algebraic sample point is chosen from every region, so the set of sample points captures all of the relevant information about the signs of the polynomials of interest \textit{for the entirety of $\mathbb{R}$}.
BKR (and Renegar) take a more indirect approach, relying on consistent sign assignments which merely indicate the \textit{existence} of points with such signs.
Consequently, although CAD will be faster in the average case, BKR and CAD have different strengths and weaknesses.
For example, CAD works best on full-dimensional decision problems~\cite{DBLP:journals/cj/McCallum93}, where only rational sample points are needed (this allows faster computation than the computationally expensive real algebraic numbers that general CAD depends on).
The Sturm-Tarski theorem is also invoked in Li \emph{et al}'s procedure to decide the sign of a univariate polynomial at a point (using only rational arithmetic)~\cite[Section 5]{li2019deciding}.
(This was later extended to bivariate polynomials by Li and Paulson \cite{li2016modular}.)
This is theoretically similar to our procedure to find the consistent sign assignments for $q_1, \dots, q_n$ at the roots of $p$, as both rely on the mathematical properties of Tarski queries; however, for example, we do not require isolating the real roots of $p$ within intervals, whereas such isolation predicates their computations.
This difference reflects our different goals: theirs is to encode algebraic numbers in \Isabelle, ours is to perform full sign determination with BKR.


PVS's $\mathtt{tarski}$ uses Tarski queries and a version of the matrix equation to solve univariate decision problems \cite{NASATarski}.
Unlike our work, $\mathtt{tarski}$ has already been optimized in significant ways; for example, $\mathtt{tarski}$ computes Tarski queries with pseudo-divisions.
However, $\mathtt{tarski}$ does not maintain a \textit{reduced} matrix equation as our work does.
Further, $\mathtt{tarski}$ was designed to solve existential conjunctive formulas, requiring DNF transformations otherwise \cite{NASATechRept}.

In addition, as previously mentioned, our work is somewhat similar in flavor to Cohen and Mahboubi's (multivariate) formalization of Tarski's algorithm \cite{AssiaQE}.
In particular, the characterization of the matrix equation and the parts of the construction that do not involve reduction share considerable overlap, as BKR derives the idea of the matrix equation from Tarski \cite{DBLP:journals/jcss/Ben-OrKR86}.
However, the reduction step is only present in BKR and is a distinguishing feature of our work.

\section{Conclusion and Future Work}\label{sec:future}
This paper describes how we have verified the correctness of a decision procedure for univariate real arithmetic in \Isabelle.
To the best of our knowledge, this is the first formalization of BKR's key insight~\cite{DBLP:journals/jcss/Ben-OrKR86,DBLP:journals/jsc/Renegar92b} for recursively exploiting the matrix equation.
Our formalization lays the groundwork for several future directions, including:
\begin{enumerate}
\item Optimizing the current formalization and adding parallelism.
\item Proving that the univariate sign determination problem is decidable in NC~\cite{DBLP:journals/jcss/Ben-OrKR86,DBLP:journals/jsc/Renegar92b} and other complexity-theoretic results. This (ambitious) project would require developing a complexity framework that is compatible with all of the libraries we use.
\item Verifying a multivariate sign determination algorithm and decision procedure based on BKR.
As mentioned previously, multivariate BKR has an error in its complexity analysis; variants of decision procedures for FOL$_{\mathbb{R}}$ based on BKR's insight that attempt to mitigate this error could eventually be formalized for useful points of comparison.
Two of particular interest are that of Renegar~\cite{DBLP:journals/jsc/Renegar92b}, who develops a full QE algorithm, and that of Canny~\cite{1993Improved}, in which coefficients can involve some more general terms, like transcendental functions.
\end{enumerate}

\section*{Acknowledgments}
We would very much like to thank Brandon Bohrer, Fabian Immler, and Wenda Li for useful discussions about Isabelle/HOL and its libraries.  Thank you also to Magnus Myreen for useful feedback on the paper.

This material is based upon work supported by the National Science Foundation Graduate Research Fellowship under Grants Nos. DGE1252522 and DGE1745016.
Any opinions, findings, and conclusions or recommendations expressed in this material are those of the authors and do not necessarily reflect the views of the National Science Foundation.
This research was also sponsored by the National Science Foundation under Grant No. CNS-1739629, the AFOSR under grant number FA9550-16-1-0288, and A*STAR, Singapore.
The views and conclusions contained in this document are those of the authors and should not be interpreted as representing the official policies, either expressed or implied, of any sponsoring institution, the U.S. government or any other entity.



\let\oldbibliography\thebibliography
\renewcommand{\thebibliography}[1]{%
  \oldbibliography{#1}%
  \addtolength{\itemsep}{-6pt}%
}

\bibliography{BKR}

\begin{thebibliography}{10}

\bibitem{algRAG}
Saugata Basu, Richard Pollack, and Marie-Fran\c{c}oise Roy.
\newblock {\em Algorithms in Real Algebraic Geometry}.
\newblock Springer, Berlin, Heidelberg, second edition, 2006.
\newblock \href {https://doi.org/10.1007/3-540-33099-2}
  {\path{doi:10.1007/3-540-33099-2}}.

\bibitem{DBLP:journals/jcss/Ben-OrKR86}
Michael Ben{-}Or, Dexter Kozen, and John~H. Reif.
\newblock The complexity of elementary algebra and geometry.
\newblock {\em J. Comput. Syst. Sci.}, 32(2):251--264, 1986.
\newblock \href {https://doi.org/10.1016/0022-0000(86)90029-2}
  {\path{doi:10.1016/0022-0000(86)90029-2}}.

\bibitem{1993Improved}
John~F. Canny.
\newblock Improved algorithms for sign determination and existential quantifier
  elimination.
\newblock {\em Comput. J.}, 36(5):409--418, 1993.
\newblock \href {https://doi.org/10.1093/comjnl/36.5.409}
  {\path{doi:10.1093/comjnl/36.5.409}}.

\bibitem{cohen_phd}
Cyril Cohen.
\newblock {\em {Formalized algebraic numbers: construction and first-order
  theory}}.
\newblock PhD thesis, {\'E}cole polytechnique, {Nov} 2012.
\newblock URL: \url{https://perso.crans.org/cohen/papers/thesis.pdf}.

\bibitem{AssiaQE}
Cyril Cohen and Assia Mahboubi.
\newblock Formal proofs in real algebraic geometry: from ordered fields to
  quantifier elimination.
\newblock {\em Log. Methods Comput. Sci.}, 8(1), 2012.
\newblock \href {https://doi.org/10.2168/LMCS-8(1:2)2012}
  {\path{doi:10.2168/LMCS-8(1:2)2012}}.

\bibitem{Collins}
George~E. Collins.
\newblock Quantifier elimination for real closed fields by cylindrical
  algebraic decomposition.
\newblock In H.~Barkhage, editor, {\em Automata Theory and Formal Languages},
  volume~33 of {\em LNCS}, pages 134--183. Springer, 1975.
\newblock \href {https://doi.org/10.1007/3-540-07407-4_17}
  {\path{doi:10.1007/3-540-07407-4_17}}.

\bibitem{NASATechRept}
Katherine Cordwell, C\'esar Mu{\~{n}}oz, and Aaron Dutle.
\newblock Improving automated strategies for univariate quantifier elimination.
\newblock Technical Memorandum NASA/TM-20205010644, NASA, Langley Research
  Center, Hampton VA 23681-2199, USA, January 2021.
\newblock URL:
  \url{https://shemesh.larc.nasa.gov/fm/papers/NASA-TM-20205010644.pdf}.

\bibitem{BKR_AFP}
Katherine Cordwell, Yong~Kiam Tan, and André Platzer.
\newblock The {BKR} decision procedure for univariate real arithmetic.
\newblock {\em Archive of Formal Proofs}, April 2021.
\newblock \url{https://www.isa-afp.org/entries/BenOr_Kozen_Reif.html}, Formal
  proof development.

\bibitem{Rank_Nullity_Theorem-AFP}
Jose Divasón and Jesús Aransay.
\newblock Rank-nullity theorem in linear algebra.
\newblock {\em Archive of Formal Proofs}, January 2013.
\newblock \url{https://isa-afp.org/entries/Rank_Nullity_Theorem.html}, Formal
  proof development.

\bibitem{Gauss_Jordan-AFP}
Jose Divasón and Jesús Aransay.
\newblock Gauss-{J}ordan algorithm and its applications.
\newblock {\em Archive of Formal Proofs}, September 2014.
\newblock \url{https://isa-afp.org/entries/Gauss_Jordan.html}, Formal proof
  development.

\bibitem{10.1145/3018610.3018617}
Jose Divas{\'{o}}n, Sebastiaan J.~C. Joosten, Ren{\'{e}} Thiemann, and Akihisa
  Yamada.
\newblock A formalization of the {Berlekamp-Zassenhaus} factorization
  algorithm.
\newblock In Yves Bertot and Viktor Vafeiadis, editors, {\em CPP}, pages
  17--29. {ACM}, 2017.
\newblock \href {https://doi.org/10.1145/3018610.3018617}
  {\path{doi:10.1145/3018610.3018617}}.

\bibitem{ducos2000optimizations}
Lionel Ducos.
\newblock Optimizations of the subresultant algorithm.
\newblock {\em J. Pure Appl. Algebra}, 145(2):149--163, 2000.
\newblock \href {https://doi.org/10.1016/S0022-4049(98)00081-4}
  {\path{doi:10.1016/S0022-4049(98)00081-4}}.

\bibitem{hales_2017}
Thomas Hales, Mark Adams, Gertrud Bauer, Tat~Dat Dang, John Harrison, Hoang
  Le~Truong, Cezary Kaliszyk, Victor Magron, Sean McLaughlin, Tat~Thang Nguyen,
  et~al.
\newblock A formal proof of the {K}epler conjecture.
\newblock {\em Forum of Mathematics, Pi}, 5, 2017.
\newblock \href {https://doi.org/10.1017/fmp.2017.1}
  {\path{doi:10.1017/fmp.2017.1}}.

\bibitem{DBLP:conf/sfm/Harrison06}
John Harrison.
\newblock Floating-point verification using theorem proving.
\newblock In Marco Bernardo and Alessandro Cimatti, editors, {\em SFM}, volume
  3965 of {\em LNCS}, pages 211--242. Springer, 2006.
\newblock \href {https://doi.org/10.1007/11757283_8}
  {\path{doi:10.1007/11757283_8}}.

\bibitem{DBLP:conf/tphol/Harrison07}
John Harrison.
\newblock Verifying nonlinear real formulas via sums of squares.
\newblock In Klaus Schneider and Jens Brandt, editors, {\em TPHOLs}, volume
  4732 of {\em LNCS}, pages 102--118. Springer, 2007.
\newblock \href {https://doi.org/10.1007/978-3-540-74591-4_9}
  {\path{doi:10.1007/978-3-540-74591-4_9}}.

\bibitem{hoelzl09realineequalities}
Johannes H{\"o}lzl.
\newblock Proving inequalities over reals with computation in {Isabelle/HOL}.
\newblock In Gabriel~Dos Reis and Laurent Th{\'e}ry, editors, {\em PLMMS},
  pages 38--45, Munich, August 2009.

\bibitem{Sturm_Tarski-AFP}
Wenda Li.
\newblock The {S}turm-{T}arski theorem.
\newblock {\em Archive of Formal Proofs}, September 2014.
\newblock \url{https://isa-afp.org/entries/Sturm_Tarski.html}, Formal proof
  development.

\bibitem{li2019deciding}
Wenda Li, Grant~Olney Passmore, and Lawrence~C. Paulson.
\newblock Deciding univariate polynomial problems using untrusted certificates
  in {Isabelle/HOL}.
\newblock {\em J. Autom. Reason.}, 62(1):69--91, 2019.
\newblock \href {https://doi.org/10.1007/s10817-017-9424-6}
  {\path{doi:10.1007/s10817-017-9424-6}}.

\bibitem{li2016modular}
Wenda Li and Lawrence~C. Paulson.
\newblock A modular, efficient formalisation of real algebraic numbers.
\newblock In Jeremy Avigad and Adam Chlipala, editors, {\em CPP}, pages 66--75.
  {ACM}, 2016.
\newblock \href {https://doi.org/10.1145/2854065.2854074}
  {\path{doi:10.1145/2854065.2854074}}.

\bibitem{10.1145/3293880.3294092}
Wenda Li and Lawrence~C. Paulson.
\newblock Counting polynomial roots in {I}sabelle/{HOL}: a formal proof of the
  {B}udan-{F}ourier theorem.
\newblock In Assia Mahboubi and Magnus~O. Myreen, editors, {\em CPP}, pages
  52--64. {ACM}, 2019.
\newblock \href {https://doi.org/10.1145/3293880.3294092}
  {\path{doi:10.1145/3293880.3294092}}.

\bibitem{DBLP:journals/cj/McCallum93}
Scott McCallum.
\newblock Solving polynomial strict inequalities using cylindrical algebraic
  decomposition.
\newblock {\em Comput. J.}, 36(5):432--438, 1993.
\newblock \href {https://doi.org/10.1093/comjnl/36.5.432}
  {\path{doi:10.1093/comjnl/36.5.432}}.

\bibitem{harrison}
Sean McLaughlin and John Harrison.
\newblock A proof-producing decision procedure for real arithmetic.
\newblock In Robert Nieuwenhuis, editor, {\em CADE}, volume 3632 of {\em LNCS},
  pages 295--314. Springer, 2005.
\newblock \href {https://doi.org/10.1007/11532231_22}
  {\path{doi:10.1007/11532231_22}}.

\bibitem{NASAHutch}
C{\'{e}}sar~A. Mu{\~{n}}oz, Anthony~J. Narkawicz, and Aaron Dutle.
\newblock A decision procedure for univariate polynomial systems based on root
  counting and interval subdivision.
\newblock {\em J. Formaliz. Reason.}, 11(1):19--41, 2018.
\newblock \href {https://doi.org/10.6092/issn.1972-5787/8212}
  {\path{doi:10.6092/issn.1972-5787/8212}}.

\bibitem{NASATarski}
Anthony Narkawicz, C{\'{e}}sar~A. Mu{\~{n}}oz, and Aaron Dutle.
\newblock Formally-verified decision procedures for univariate polynomial
  computation based on {Sturm}'s and {Tarski}'s theorems.
\newblock {\em J. Autom. Reason.}, 54(4):285--326, 2015.
\newblock \href {https://doi.org/10.1007/s10817-015-9320-x}
  {\path{doi:10.1007/s10817-015-9320-x}}.

\bibitem{DBLP:journals/jar/Nipkow10}
Tobias Nipkow.
\newblock Linear quantifier elimination.
\newblock {\em J. Autom. Reason.}, 45(2):189--212, 2010.
\newblock \href {https://doi.org/10.1007/s10817-010-9183-0}
  {\path{doi:10.1007/s10817-010-9183-0}}.

\bibitem{DBLP:books/sp/NipkowPW02}
Tobias Nipkow, Lawrence~C. Paulson, and Markus Wenzel.
\newblock {\em Isabelle/HOL - {A} Proof Assistant for Higher-Order Logic},
  volume 2283 of {\em LNCS}.
\newblock Springer, 2002.
\newblock \href {https://doi.org/10.1007/3-540-45949-9}
  {\path{doi:10.1007/3-540-45949-9}}.

\bibitem{DBLP:journals/jar/Paulson89}
Lawrence~C. Paulson.
\newblock The foundation of a generic theorem prover.
\newblock {\em J. Autom. Reason.}, 5(3):363--397, 1989.
\newblock \href {https://doi.org/10.1007/BF00248324}
  {\path{doi:10.1007/BF00248324}}.

\bibitem{DBLP:conf/lpar/PaulsonB10}
Lawrence~C. Paulson and Jasmin~Christian Blanchette.
\newblock Three years of experience with {S}ledgehammer, a practical link
  between automatic and interactive theorem provers.
\newblock In Geoff Sutcliffe, Stephan Schulz, and Eugenia Ternovska, editors,
  {\em IWIL}, volume~2 of {\em EPiC Series in Computing}, pages 1--11.
  EasyChair, 2010.

\bibitem{Platzer18}
Andr{\'e} Platzer.
\newblock {\em Logical Foundations of Cyber-Physical Systems}.
\newblock Springer, Cham, 2018.
\newblock \href {https://doi.org/10.1007/978-3-319-63588-0}
  {\path{doi:10.1007/978-3-319-63588-0}}.

\bibitem{DBLP:conf/cade/PlatzerQR09}
Andr{\'{e}} Platzer, Jan{-}David Quesel, and Philipp R{\"{u}}mmer.
\newblock Real world verification.
\newblock In Renate~A. Schmidt, editor, {\em CADE}, volume 5663 of {\em LNCS},
  pages 485--501. Springer, 2009.
\newblock \href {https://doi.org/10.1007/978-3-642-02959-2_35}
  {\path{doi:10.1007/978-3-642-02959-2_35}}.

\bibitem{Matrix_Tensor-AFP}
T.V.H. Prathamesh.
\newblock Tensor product of matrices.
\newblock {\em Archive of Formal Proofs}, January 2016.
\newblock \url{https://isa-afp.org/entries/Matrix_Tensor.html}, Formal proof
  development.

\bibitem{DBLP:journals/jsc/Renegar92b}
James Renegar.
\newblock On the computational complexity and geometry of the first-order
  theory of the reals, part {III:} {Q}uantifier elimination.
\newblock {\em J. Symb. Comput.}, 13(3):329--352, 1992.
\newblock \href {https://doi.org/10.1016/S0747-7171(10)80005-7}
  {\path{doi:10.1016/S0747-7171(10)80005-7}}.

\bibitem{pittir16721}
Alexey Solovyev.
\newblock {\em Formal Computations and Methods}.
\newblock PhD thesis, University of Pittsburgh, Jan 2013.
\newblock URL: \url{https://d-scholarship.pitt.edu/16721/}.

\bibitem{Matrix-AFP}
Christian Sternagel and René Thiemann.
\newblock Executable matrix operations on matrices of arbitrary dimensions.
\newblock {\em Archive of Formal Proofs}, June 2010.
\newblock \url{https://isa-afp.org/entries/Matrix.html}, Formal proof
  development.

\bibitem{DBLP:journals/mics/Sturm17}
Thomas Sturm.
\newblock A survey of some methods for real quantifier elimination, decision,
  and satisfiability and their applications.
\newblock {\em Math. Comput. Sci.}, 11(3-4):483--502, 2017.
\newblock \href {https://doi.org/10.1007/s11786-017-0319-z}
  {\path{doi:10.1007/s11786-017-0319-z}}.

\bibitem{Tarski}
Alfred Tarski.
\newblock {\em A Decision Method for Elementary Algebra and Geometry}.
\newblock RAND Corporation, Santa Monica, CA, 1951.
\newblock Prepared for publication with the assistance of J.C.C. McKinsey.
\newblock URL: \url{https://www.rand.org/pubs/reports/R109.html}.

\bibitem{Jordan_Normal_Form-AFP}
René Thiemann and Akihisa Yamada.
\newblock Matrices, {Jordan} normal forms, and spectral radius theory.
\newblock {\em Archive of Formal Proofs}, August 2015.
\newblock \url{https://isa-afp.org/entries/Jordan_Normal_Form.html}, Formal
  proof development.

\bibitem{Algebraic_Numbers-AFP}
René Thiemann, Akihisa Yamada, and Sebastiaan Joosten.
\newblock Algebraic numbers in {Isabelle/HOL}.
\newblock {\em Archive of Formal Proofs}, December 2015.
\newblock \url{https://isa-afp.org/entries/Algebraic_Numbers.html}, Formal
  proof development.

\bibitem{10.1145/2854065.2854073}
Ren{\'{e}} Thiemann and Akihisa Yamada.
\newblock Formalizing {J}ordan normal forms in {Isabelle/HOL}.
\newblock In Jeremy Avigad and Adam Chlipala, editors, {\em CPP}, pages 88--99.
  {ACM}, 2016.
\newblock \href {https://doi.org/10.1145/2854065.2854073}
  {\path{doi:10.1145/2854065.2854073}}.

\bibitem{10.1145/800061.808728}
Joachim von~zur Gathen.
\newblock Parallel algorithms for algebraic problems.
\newblock {\em {SIAM} J. Comput.}, 13(4):802--824, 1984.
\newblock \href {https://doi.org/10.1137/0213050} {\path{doi:10.1137/0213050}}.

\end{thebibliography}

\appendix
\section{Comments on Multivariate BKR}\label{sec:multivar}
The ultimate intent is for the univariate formalization to serve as the basis for an extension to the multivariate case.
%
%
The main part of the univariate construction that must be adapted for multivariate polynomials is the computation of Tarski queries.
In the univariate case, this is accomplished with remainder sequences per the following (standard) result:

\begin{theorem}\label{thm:Sturm}[Generalized Sturm's theorem~\cite[Proposition 8.1]{DBLP:journals/jsc/Renegar92b}] Given coprime univariate polynomials $p$, $q$ with $p \neq 0$, form the Euclidean remainder sequence $p_1 = p$, $p_2 = p'q$, and $p_i$ is the negated remainder of $p_{i-2}$ divided by $p_{i-1}$ for $i \geq 3$.
This terminates at some $p_{k+1} = 0$ because the remainder has lower degree than the divisor at every step.
Let $a_i$ be the leading coefficient of $p_i$ for $1 \leq i \leq k$.
Consider the two sequences $a_1, \dots, a_k$ and $(-1)^{\deg p_1}a_1, \cdots, (-1)^{\deg p_k}a_k$.
If $S^+(p, q)$ is the number of sign changes in $a_1, \dots, a_k$ and $S^-(p, q)$ is the number of sign changes in $(-1)^{\deg p_1}a_1, \cdots, (-1)^{\deg p_k}a_k$, then $N(p, q) = S^-(p, q) - S^+(p, q)$.
\end{theorem}

Following the idea of BKR, we intend to treat multivariate polynomials in $n$ variables as univariate polynomials (whose coefficients are polynomials in $n-1$ variables) and so compute remainder sequences of polynomials with attention to a single variable.
These remainder sequences will be sequences of polynomials in $n-1$ variables rather than integers, but we only need to know the \emph{signs} of those polynomials (rather than their values).
That reduces the problem of sign determination for polynomials in $n$ variables to a sign determination problem for polynomials in $n-1$ variables.
In this way we intend to successively reduce multivariate computations to a series of (already formalized) univariate computations.

This intuition can be captured by the following concrete example.
Consider $p = x^2y + 1$ and $q = xy + 1$.
Suppose we choose to first eliminate $y$.
If $x$ is 0, then the analysis for the remaining $p = q = 1$ is simple.
Otherwise, both $x$ and $x^2$ are nonzero.
Now, we calculate the remainder sequence from \rref{thm:Sturm}:
$p_1 = x^2y + 1$, $p_2 =  x^3y + x^2$, and $p_3 = -(1-x)$.
To find $p_3$, we calculate $x^2y + 1 = \frac{1}{x}(x^3y + x^2) + (1 - x)$, where $\frac{1}{x}$ is well-defined since $x \neq 0$.

The leading coefficients of $p_1, p_2,$ and $p_3$ as polynomials in $y$ are $a_1 = x^2$, $a_2 = x^3$, and $a_3 = -(1-x)$.
Here, we must use our univariate algorithm to fix some consistent sign assignment in $x$ on the $a_i$'s, taking into account our earlier stipulation that $x$ and $x^2$ are nonzero.
Say that we choose, for example, $x$ positive, $x^3$ positive, and $-(1-x)$ negative.
(A full QE procedure would need to consider \emph{all} possible consistent sign assignments.)
Because of our chosen sign assignment, $a_1$ is positive, $a_2$ is positive, and $a_3$ is negative. Still following \rref{thm:Sturm}, $S^+(p, q) = 1$ and $S^-(p, q) = 0$.
The Tarski query $N(\{1\})$ is then computed as $N(\{1\}) = N(p, q) = S^-(p, q) - S^+(p, q) = -1$.

If we wish to find the signs of $q$ at the roots of $p$, we can use this way of computing Tarski queries to build the matrix equation for $p$ and $q$. Computing $N(\{\}) = 1$, and following the method of the base case (in which the candidate signs list is $[[+1],[-1]]$), we find:
$$\begin{pmatrix} 1 & 1\\ 1 & -1 \end{pmatrix} \begin{pmatrix} 0 \\ 1 \end{pmatrix} = \begin{pmatrix} 1 \\ -1 \end{pmatrix}$$
Looking at the LHS vector, we see that only its second entry is nonzero.
This means that $-1$ is the only consistent sign assignment for $q$ at the zeros of $p$, \textit{given our assumptions that $x$ is positive and $-(1 - x)$ is negative}.

We can check this as follows: Given our assumption that $x \neq 0$, the only root of $p$ is $-\frac{1}{x^2}$.
Plugging this into $q$, we obtain $x(-\frac{1}{x^2}) + 1 = -\frac{1}{x} + 1$. Because $x$ is assumed to be positive, the sign of $-\frac{1}{x} + 1$ is the same as the sign of $x(-\frac{1}{x} + 1) = -1 + x = -(1-x)$, which we have assumed to be negative.

Thus, $-1$ is a consistent sign assignment for $q$ at the roots of $p$.
To find the other consistent sign assignments, we repeat this process with all other consistent choices for the signs of $x$ and $a_1, a_2, a_3$.

\end{document}